\documentclass[preprint2]{aastex}

\slugcomment{Accepted March 19, 2015 for publication in the Astrophys. J.}

\shorttitle{Variability in PPNs: III. LMC/SMC C-rich Objects}

\shortauthors{Hrivnak et al.}

\begin{document}


\title{VARIABILITY IN PROTO-PLANETARY NEBULAE: III. LIGHT CURVE STUDIES OF MAGELLANIC CLOUD CARBON-RICH OBJECTS}


\author{Bruce J. Hrivnak\altaffilmark{1}, Wenxian Lu\altaffilmark{1}, Kevin Volk\altaffilmark{2}, Ryszard Szczerba\altaffilmark{3}, Igor Soszy\'{n}ski\altaffilmark{4}, and Marcin Hajduk\altaffilmark{3}}

\altaffiltext{1}{Department of Physics and Astronomy, Valparaiso University,
Valparaiso, IN 46383; bruce.hrivnak@valpo.edu. wen.lu@valpo.edu}
\altaffiltext{2}{Space Telescope Science Institute, 3700 San Martin Drive, Baltimore, Maryland, 21218, USA}
\altaffiltext{3}{N. Copernicus Astronomy Center, Rabianska 8, 87-100 Torun, Poland }
\altaffiltext{4}{Warsaw University Observatory, Al. Ujazdowskie 4, 00-478 Warsaw, Poland}

\begin{abstract}

We have investigated the light variability in a sample of 22 carbon-rich post-AGB stars in the Large Magellanic Cloud (LMC) and Small Magellanic Cloud (SMC), based primarily on photometric data from the OGLE survey.
All are found to vary.
Dominant periods are found in eight of them;  
these periods range from 49 to 157 days, and most of these stars have F spectral types. 
These eight are found to be similar to the Milky Way Galaxy (MWG) carbon-rich proto-planetary nebulae (PPNs) in several ways:
(a) they are in the same period range of $\sim$38 to $\sim$160 days, 
(b) they have similar spectral types,
(c) they are (all but one) redder when fainter, 
(d) they have multiple periods, closely spaced in time, with a average ratio of secondary to primary period of $\sim$1.0,
and as an ensemble,
(e) they show a trend of decreasing period with increasing temperature, and
(f) they show a trend of decreasing amplitude with decreasing period.
However, they possibly differ in that the decreasing trend of period with temperature may be slightly offset from that of the MWG.
These eight are classified as PPNs.
The other 14 all show evidence of variability on shorter timescales.  They are likely hotter PPNs or young planetary nebulae.
However, in the MWG the numbers of PPNs peak in the F$-$G spectral types, while it appears that in the LMC they peak at a hotter B spectral type.
One of the periodic ones shows a small, R Coronae Borealis-type light curve drop.

\end{abstract}




\section{INTRODUCTION}
\label{intro}

Proto-planetary nebulae (PPNs) are objects in transition between the asymptotic giant branch (AGB) and planetary nebula (PN) phases in the evolution of intermediate- and low-mass stars.
As such, they are characterized by a central star of spectral type B$-$G surrounded by an expanding envelope of gas and dust.\footnote{Note that PPNs are a subset of post-AGB objects, since not all post-AGB objects will evolve at the appropriate rate to photoionize a detached, expending shell before it disperses.  RV Tauri stars appear to be an example of post-AGB stars that will not evolve into PNs.}  
Selecting PPN candidates is difficult using only visible light observations, but the excess mid-infrared emission produced by their cool dust affords a good means to identify PPN candidates. 
Thus, following the {\it IRAS} mission, it became possible to identify a number of candidates and to confirm them with ground-based observations \citep{pot88,hri89}, followed up with {\it HST} imaging \citep{ueta00,su01} and {\it ISO} \citep{hri00} and {\it Spitzer} \citep{hri09} mid-infrared spectroscopy.
It was seen that most PPNs could be characterized as oxygen-rich (C/O$<$1) or carbon-rich (C/O$>$1) based on mid-infrared spectral features.  
The O-rich sources display silicate emission or absorption features at 9.7 and 18 $\mu$m and some show crystalline silicate emission features at somewhat longer wavelengths.
The less common C-rich sources show unidentified infrared features (UIRs) at 6.2, 7.6, 8.6 and 11.3 $\mu$m commonly attributed to polycyclic aromatic hydrocarbons, with a broad feature around 30 $\mu$m (the so-called ``30 $\mu$m feature'') often attributed to MgS and an as yet unidentified feature at 20 $\mu$m (the so-called ``21 $\mu$m feature''). 

AGB and post-AGB objects, such as RV Tauri variables, are known to pulsate, so it was reasonable to assume that the central stars in PPNs might also vary in brightness due to pulsation. 
Such variability has indeed been seen in Galactic PPNe \citep{ark00,ark10,ark11,hri01,hri10}.  
\citet{hri10} have published the results of a long-term (14-year) photometric study of twelve Galactic C-rich PPNs.  They are of spectral types F and G and all show the 21 $\mu$m feature.  Periodic variability is found in each of them, with periods ranging from 
38 to 153 days.
In addition,  a clear linear correlation was found to exist between both pulsation period and effective temperature and between pulsation period and pulsation amplitude over this F$-$G spectral range.  
In still hotter PPNs, shorter-term variability is seen \citep[and references therein]{ark06}. 

A good sample of 17 candidates for C-rich PPNs in the LMC and SMC has been identified by
\citet{volk11}, on the basis of {\it Spitzer} mid-infrared spectra.
They all have strong mid-infrared excesses with dust temperatures of 200$-$350 K.
They show a range of emission features characteristic of C-rich objects $-$ UIRs, 21 $\mu$m feature, 30 $\mu$m feature, and perhaps SiC in a few $-$ although most show only some of these features.
Nine show the 21 $\mu$m feature and in four others it is either present but very weak or of questionable presence.
In a related study, \citet{sloan14} enlarged this sample with an additional 25 post-AGB objects in the LMC/SMC that are identified as C-rich on the basis of their {\it Spitzer} mid-infrared spectra. Twenty-two of these have been classified as PNs.
Many of these objects have previously been mistakenly classified as young stellar objects (YSOs) or YSO candidates based primarily upon their infrared colors.  Indeed, in most color-color and color-magnitude diagrams made from near-infrared and mid-infrared magnitudes, there is a strong overlap between YSOs and post-AGB objects. 
The detailed analysis by \citet{volk11}, which included modeling of the double-peaked SEDs, the published visible-band spectroscopic analysis of several of these as discussed later in this paper, and our variability study shows that these objects are PPNs rather than YSOs.

In this paper, we present the results of a study of the light variability of C-rich LMC and SMC PPN candidates, based primarily on light curves from the Optical Gravitational Lensing Experiment (OGLE) project.
The PPN candidates are primarily the 17 identified by \citet{volk11}.  
The sample was increased slightly by the addition of the three non-PN sources included in the study by \citet{sloan14}.
Finally, we included two additional LMC objects studied by \citet{aarle11} that are shown to be C-rich based on high-resolution abundance analyses \citep{aarle13}.
Unfortunately, neither of these later two has a {\it Spitzer} mid-infrared spectrum.
However, both have large infrared excesses consistent with PPNs with detached circumstellar shells, and both have infrared colors, K$-$[24] versus K$-$[8.0], consistent with the colors of known PPN candidates,
where the K magnitudes are from the 2MASS survey and the [8.0] and [24] magnitudes are from the Spitzer Space Telescope observations.
Our goal is to investigate the photometric variability in this sample of C-rich PPN candidates to determine periods and amplitudes of pulsation where they exist, and to compare them to the relationships found for C-rich PPNs in the MWG.
Since the metallicities are different between the LMC, SMC, and MWG, this also affords the potential opportunity to investigate differences in pulsation properties due to composition.  
We will also attempt to classify the nature of each object.
The 22 objects are listed in Table~\ref{object_list}.
For convenience, we will often list them by their number in Table~\ref{object_list}.

\placetable{object_list}

\section{DATA SETS AND ANALYSIS METHODS}
\label{data}

The main data sets that we used in this study are from the OGLE surveys, OGLE-II \citep{szy05} and OGLE-III\footnote{The OGLE homepage is found at http://ogle.astrouw.edu.pl/} \citep{uda08}.  For some of the objects the data are publicly available and for others they are still proprietary.  The OGLE-II data set covers the years 1997$-$2000 and involves five seasons of observing in the LMC and four seasons of observing in the SMC, with most of the data with the Cousins {\it I} filter\footnote{In keeping with OGLE practice, we have indicated the Cousins {\it I} without including a subscript.}
 but some with the Johnson {\it V} filter.  The OGLE-III data set covers the years 2001-2009 in the {\it I} filter, involving eight seasons of observing, with occasional observations in the {\it V} filter.  For some of the fainter sources, we have removed data points with unusually large uncertainties.  We formed ({\it V$-$I}) color data points from {\it V} and {\it I} observations that were made on the same night.  OGLE-III data are available for 19 of the 22 objects (all but objects 5, 8, and 14) and OGLE-II for five  objects (2, 4, 9, 12, and 18). 
For the three objects without OGLE data (5, 8, and 14), we used data from the MACHO\footnote{The MACHO homepage is found at http://wwwmacho.anu.edu.au/} Project \citep{alc97}. 
These cover the years from late-1992/early-1993 to late-1999/early-2000 (depending upon the object).
The magnitudes listed in the MACHO catalog are instrumental magnitudes.
To give more meaningful values for the graphs that we display for these objects, we adjusted them to  approximately standard magnitudes using zero-point offsets derived from published photometric measurements for these objects.  To designate that they are only approximate, non-standard values, 
we list them with primes: {\rm V$^{\prime}$} and {\rm R}$^\prime_C$.  
The subsequent analyses were done with the original, instrumental MACHO catalog data.
In Table~\ref{statistics} are listed some statistics of the data sets used, including the average uncertainties of the data.
The average uncertainties in the observations range from $\sim$0.005 mag for the brighter objects ({\it I}=13-15 mag, {\it V}$\approx$15.5 mag) to $\sim$0.1 mag ({\it I}) and $\sim$0.2 mag ({\it V}) for the fainter objects ({\it I}=19-20 mag, {\it V}=21-22 mag).

\placetable{statistics}

In Figure~\ref{fig1} are shown the {\it I} light curves of the five 
objects with both OGLE-II and OGLE-III data and in Figures~\ref{fig2a} and \ref{fig2b} the 
objects with only OGLE-III data. 
In Figure~\ref{fig3} are shown the {\rm R}$^\prime_C$ light curves of the three objects with only MACHO data.
The OGLE {\it V} versus {\it V$-$I} and the MACHO {\rm V$^{\prime}$} versus {\rm V$^{\prime}$}$-${\rm R}$^\prime_C$ color curves are shown in Figure~\ref{color}. 
For three of the objects (14, 17, 19) there are no {\it V} observations and thus no color curves.

\placefigure{fig1}

\placefigure{fig2a}

\placefigure{fig2b}

\placefigure{fig3}

\placefigure{color}

The light curves were were first examined visually to investigate variability within a season and also long-term variability.  All are seen to vary.
They were then analyzed for periodicity.
We used the period search programs Period04 \citep{lenz05} and CLEAN \citep{rob87}, both of which are based on a Fourier transform. The results of the frequency analyses were the same.
We present the results from Period04 because it is more versatile, since it allows one to determine the signal-to-noise ratio (S/N) of a given period, fits the light curve with a sine curve of calculated amplitude, and can easily be used to investigate multiple periods.     
To judge if a period was significant, we used the recommended Period04 criterion that if the S/N amplitude ratio of the peak in the frequency power spectrum was $\ge$ 4, it was regarded as significant \citep{bre93}.
For those with significant periods of variation, we also explored the existence of multiple periods.
Some objects show long-term seasonal trends in the light level or what appear to be season-to-season variations in the mean light levels.
For the former of these, we first removed the long-term trends before conducting the period analysis. 
For those that showed the season-to-season variations, we analyzed the light curves as observed and also as normalized to the seasonal average values.
This normalization was done by calculating the average value within a season, adjusting the seasonal averages to a common value (0.00), and then doing the period study.  We typically determined similar periods for the observed and seasonally-normalize data.
In a few cases a long period ($>$ 1 year) was suggested that modulated the seasonal light curve levels.
Since our goal was to determine pulsational periods, we have chosen to use the results from the normalized light curves.
The question of the cause of these longer-term variations will be addressed in Section~\ref{discussion}.
The analyses of the OGLE data were carried out 
on the {\it I} light curves and the analyses of the MACHO data were carried out on both the instrumental 
{\rm V$^{\prime}$} and {\rm R}$^\prime_C$ light curves.

\section{PHOTOMETRIC VARIABILITY OF THE INDIVIDUAL OBJECTS}
\label{Variability}

All of the objects are seen to vary in light.  Only eight of them show clearly periodic variability, with one other showing less certain periodicity.  These nine are discussed individually in this section; the others are discussed in Appendix I.
These objects have been included in large surveys of variable stars in the LMC or SMC \citep{ita04, fraser08, sos09, sos11, spano11}, 
with periods cited for 10 of them. 
In most but not all cases, the periods determined are similar to ours.  
We will compare the results of our period study with others where they exist.   
\citet{fraser08} analyzed the MACHO data to look for periods.
\citet{sos09,sos11} used the OGLE-III data, although we will refer instead to the more detailed results available in the 
OGLE-III Catalog of Variable Stars.\footnote{The OGLE-III On-line Catalog of Variable stars is found at http://ogledb.astrouw.edu.pl/~ogle/CVS/} 
\citet{spano11} carried out a search for long-period variables based on the EROS-2 survey of the LMC, with a mean observing duration of $\sim$6 years, and listed periods for three of our objects, although for only one of them did we find a period.  
Observations were carried out between 1996 and 2003, using blue and red filters; more details of the EROS-2 survey are described in their paper.
\citet{ita04} searched for dust-enshrouded variables in the LMC and SMC, and found the period for one object in common with our study.

(1) {\it J003659.53-741950.4} (Figure~\ref{fig2a}):
The object is very bright, with $<${\it I}$>$ = 13.04,  
and is red, with $<${\it V$-$I}$>$ = 1.61.
The ${\it I}$ data show a clear variation in brightness, with a peak-to-peak range of 0.21 mag over these eight years of observations.  The $<${\it I}$>$ light curves show a season-to-season range in amplitude of 0.12 (2007) to 0.21 (2004) mag, with an average variation of 0.16 mag in a season.  The mean light level also changes slightly from season to season over a range of 0.04 mag, but without any general trends.
Visual inspection of the light curve shows a cyclical behavior with cycle lengths ranging from $\sim$50 (2007) to $>$150 (2002) days.  
The period analysis of the data as observed and also as normalized yielded the same result, with a primary period of 142.4$\pm$0.2 days, a slightly larger-amplitude period of 356 days, and a third period of 115.6 days for the adjusted light curve.
The power spectrum and phase diagram for P$_1$ are shown in Figure~\ref{Pspec}.
The second period is suspiciously close to one year, so it might be an alias due to the sampling.  But the other two periods are robust. 
The light curve fitted with these three periods is displayed in Figure~\ref{obj20n_LC}.  The fit is reasonably good.    
These period values are significantly different than that found in the OGLE-III Variable Star Catalog, where it is listed as long-period variable with a period 102.0 days, with additional periods of 356 and 231 days.
The ${\it V}$ data appear to follow the ${\it I}$ light curve, but are many fewer in number and are only for the last five years.
The color variation is small, 0.07 mag, and is not correlated with brightness.

\placefigure{obj20n_LC}

\begin{figure}\figurenum{1}\epsscale{1.10} 
\plotone{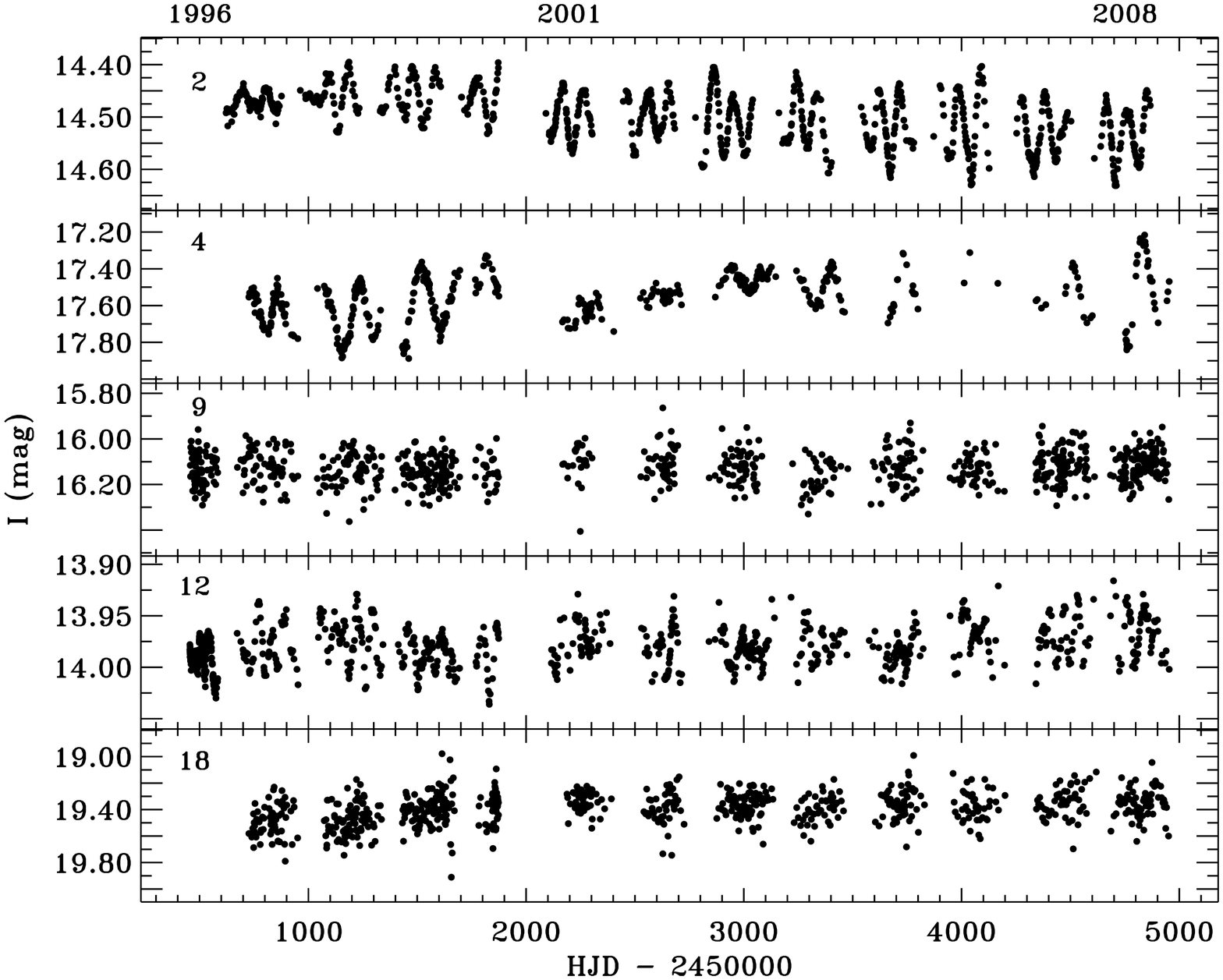}
\caption{The {\it I} light curves for the five objects with both OGLE-II and OGLE-III data, from 1996 to 2008.  The numbers in the upper left of each panel refer to the object identifications listed in Table~\ref{object_list}.
\label{fig1}}
\epsscale{1.0}
\end{figure}

(2)   {\it J004441.03-732136.44} (Figure~\ref{fig1}):
This object has data from both OGLE-II and OGLE-III and the two data sets were combined for an overall analysis.  This object was observed intensively over 12 seasons, with a total of 1028  observations in {\it I}.  It is bright, $<${\it I}$>$ = 14.50 and the average uncertainty in the observations is small $<$$\sigma$$_ I$$>$ = 0.004.  The observations show a clear variability, with a typical variation of 0.15 mag in a season and an overall 
variation of 0.23 mag ({\it I}).  The 12-year light curve appears to reach a maximum in overall brightness in 1998-1999 and then decrease by 0.08 mag by 2007-2008.  An examination of the data shows a very clear cyclical pattern with a modulated or beat amplitude, with a peak-to-peak amplitude of 0.05 mag in the first season increasing to 0.13 mag in the second and a maximum 0.22 mag in 2006.  A period analysis of the seasonally-normalized data reveals a dominant period of 96.3 $\pm$0.1 days.  
The frequency power spectrum and the data phased to this period are shown in Figure~\ref{Pspec}.
There also exist secondary periods at 99.6 and 94.4 days of similar, lower amplitude that serve to modulate the overall amplitude and slightly shift the maxima and minima to better fit the observed light curve.
In Table~\ref{periods} are listed these periods, with their associated amplitudes and phases.
The light curve is fitted very well by these three periods, as is shown in Figure~\ref{obj1_LC}.
Similar values are found when examining only the OGLE-III data set (see Table~\ref{periods}).
It has been classified as a long-period, semi-regular variable (SMC$-$LPV$-$04910) in the OGLE-III Catalog of Variable Stars
 \citep[see][]{sos11}, with periods identical to those we found.  
\citet{smedt12} determined a similar dominant period of P = 97.6 $\pm$0.3 days based only on the four years of OGLE-II data.
A smaller number  (82) of {\it V} observations were made, mostly over the last five years.  These show a larger range of variation of 0.40 mag, and a clear cyclical pattern is seen even in this small number of  observations.  Similarly a cyclical variation is seen in the ({\it V$-$I}) color, with a range of 0.18 mag based on 52 observations.  Formal periodogram analyses of the {\it V} and the ({\it V$-$I}) data each indicate periods of 96.7 and 96.2 days, 
respectively, consistent with the period based on the much more numerous {\it I} observations.  The object is red, with $<${\it V$-$I}$>$ = 1.59, and shows a strong correlation of brightness with color, being redder when fainter (see Figure~\ref{color}). \\

\placefigure{Pspec}

\placetable{periods}

\placefigure{obj1_LC}

(4) {\it J050603.66-690358.9} (Figure~\ref{fig1}):
This object is medium bright and very red, with $<${\it I}$>$ = 17.56 mag and $<${\it V$-$I}$>$ = 3.4.  
The light curve shows changes in both the seasonal amplitude and seasonal average brightness.
The OGLE-II data from 1997$-$2000 show an increasing amplitude for the first three seasons and in general an increasing average brightness.  This is followed by what appears to be a discontinuity between the OGLE-II and III data, although this may represent a real change in the average brightness.
The OGLE-III data from 2001$-$2008 start at about the average level of the 1997 data and the light curve shows a general trend of increasing in average brightness from 2001 to 2003 and then is relatively constant in average brightness 2003$-$2008.  The range of variability increases markedly from 0.13 mag in 2002 to 0.62 mag in 2008.  
Inspection of the data in the individual seasons shows a clear cyclical variation except in the years 2001 and especially 2002. 

 An analysis was carried out of the seasonally-normalized data, with season 10 (2006) removed since it had only three observations.  
 A fit to all of the data resulted in three periods, a dominant period of P$_1$ = 157.2 $\pm$ 0.1 days and two additional periods 137.9 and 239.0 days of similar amplitudes.
 The frequency power spectrum and the data phased with P$_1$ are shown in Figure~\ref{Pspec}.  In Figure~\ref{obj3_LC} is plotted the light curve fitted with the three periods and their amplitudes and phases, which are listed in Table~\ref{periods}.
The fit is reasonably good, except for 2001 and 2002.
We also analyzed the {\it I} light curve without the data from 2001 and 2002.
and the results were similar to what was found for the full data set.
 The OGLE-III {\it V} light curve varies over a range of 0.9 mag and the {\it V$-$I} color curve over a range of 0.6 mag, with most of the data and most of the variation found in the last season.  There are no reliable {\it V} data from the OGLE-II study.  There is a general correlation between color and brightness, in this case with a large dispersion, and the object is redder when fainter (Figure~\ref{color}).
This object is not listed in the OGLE-III Catalog of Variable Stars.  
However, it is listed in the study of variable stars in the Magellanic Clouds by \citet{ita04}, who found a relatively similar dominant period of 159.49 days based on OGLE-II data. \\
 
\placefigure{obj3_LC}

(5) {\it J050632.10-714229.8} (Figure~\ref{fig3}): This object was not observed in the OGLE surveys but was observed in the MACHO survey.  It is bright, $<${\it I}$>$ = 13.78 and relatively blue, 
$<${\it V$-$I}$>$ = 0.52 \citep{zar97}.
Both the {\rm V$^\prime$} and {\rm R}$^\prime_C$ light curves show that the object varies in brightness, with a slight long-term decreases in brightness over the observing interval of 0.04 ({\rm V$^{\prime}$}) and 0.03 ({\rm R}$^\prime_C$) mag.  
They each show some suggestion of cyclical variability.  
Neglecting a few outliers, the {\rm R}$^\prime_C$ light curve varies over a range of 0.15 mag peak-to-peak and the {\rm V$^{\prime}$} light curve over a range of 0.22 mag.
With the decreasing trend in brightness first removed, both the {\rm R}$^\prime_C$ and {\rm V$^{\prime}$} light curves were analyzed.
Both indicate a period of 49.0 days, but in the {\rm R}$^\prime_C$ light curve this is just below the level of significance while in the {\rm V$^{\prime}$} light curve it is significant. 
This is almost the same period of 49.10 days found by \citet{aarle11}, who classified it as a population II Cepheid.
Further analysis of the {\rm V$^{\prime}$} light curve results in two periods, 48.9$\pm$ 0.1 and 53.5$\pm$ 0.1 days.  The frequency spectrum and the phased light curve are shown in Figure~\ref{Pspec}.
In Figure~\ref{obj19_LC} is shown the {\rm V$^{\prime}$} light curve fitted with these two periods.  The fit is of fair quality, not as good as that found for most of the objects in this study.
The object shows a clear trend of being redder when fainter (Figure~\ref{color}).
 
\placefigure{obj19_LC}

 (8) {\it J051110.64-661253.7} (Figure~\ref{fig3}): This object was not observed in the OGLE surveys; however it does have MACHO data.  It is bright, $<${\it I}$>$ = 14.54 and red, $<${\it V$-$I}$>$ = 2.09 \citep{zar97}. 
 The {\it R$_C$}$^\prime$ light curve shows clear evidence of variability, with a peak-to-peak variation of 0.14 mag and seasonal variations that range from about 0.07 to 0.14 mag.  The average brightness appears to be about the same except for the last season, which is about 0.02 mag fainter.  
We normalized the light curve by adjusting the last year to be at the mean level of the other years. 
Visual inspection of the light curves show cyclical variations of $\sim$120 days.  The period analysis indicates two significant periods,  111.8 and 95.7 days, with the longer one slightly stronger.  
These are listed in Table~\ref{periods}.  The frequency spectrum and the phase plot are shown in Figure~\ref{Pspec}.  In Figure~\ref{obj6_LC} is shown the fit of these two periods.
The light curve is complex, and the fit is only of fair quality.  It is not as good a fit as found most of the other objects for which we find periods.  The V$^\prime$ light curve looks similar, with a slightly larger peak-to-peak variation of 0.19 mag.  An analysis of the V$^\prime$ light curve results in similar periods of 96.2 and 112.0 days, along with a low-amplitude (semi-amplitude = 0.018 mag), longer period of 500 days.
However, for the V$^\prime$ light curve, the but the shorter period one slightly stronger. 
Thus the two periods appear to be approximately equally strong, with the difference found between the two bandpasses likely resulting from the slightly different time sampling (the V$^\prime$ light curve has 49 more data points).
Even the (V$^\prime$$-$R$_{\rm C}$$^\prime$) color curves gives an indication of a similar period of 112 days, but at a S/N that is somewhat below the level to be regarded as significant.
\citet{fraser08} found similar periods of 96.4 and 111.6 days, while \citet{aarle11} classified it as a semi-regular variable but did not list a period.  The fact that \citet{fraser08} found the same two periods based on a different data set (MACHO) increases our confidence in the results in spite of the inferior fit to the light curve.
There is a clear correlation between color and brightness (Figure~\ref{color}).\\
 
\placefigure{obj6_LC}
 
(12) {\it J052043.86-692341.0 (LMC100.2\_18930)} (Figure~\ref{fig1}):
This object is very bright: $<${\it I}$>$ = 13.98 
and $<${\it V$-$I}$>$ = 1.22.  It was observed in both the OGLE-II and OGLE-III surveys, and there are 13 seasons of observations with 957 {\it I} data points.  The entire range of variability is small, only 0.12 mag peak-to-peak, with a typical variation of only 0.07-0.08 mag ({\it I}) in a season.  The seasonal mean and maximum brightnesses both appear to show some small variation over a range of 0.03 mag.  An examination of the data season by season shows a cyclical variation in most seasons.  This is especially seen in the first season (1997), which has 257 data points.  For the period analysis, we normalized the data set to the seasonal average values to remove longer-term seasonal variations.  A period study of the entire data set reveals several significant periods of 
74.2, 79.9, 57.7, and 76.9 days.  In Figure~\ref{Pspec} is shown the frequency spectrum for the {\it I} data and the phase plot based on P$_1$ = 74.2$\pm$0.1 days.
Together these four periods give a reasonable fit to the light curve maxima and minima, although they do not give a good fit to the first season of data.  This is shown in Figure~\ref{obj10_LC}. 
We also investigated the data in smaller time intervals of one or two seasons; the most common periods have values in the range 66 to 79 days, in agreement with the values found from the entire data set. On the first season, with a very well-defined light curve of 257 data points over an interval of 128 days, significant periods of 47 and 38 days were found.  So the variability, although cyclical in appearance, is complex.  It is listed as a long-period variable (LMC$-$LPV$-$43685) in the OGLE-III Catalog of Variable Stars, with P$_1$ = 74.15, P$_2$ = 80.12, and P$_3$ = 57.75 d, in agreement with what we found for the combined OGLE-II and III data sets.  
\citet{fraser08} find a similar period of 74 days based on the MACHO data set.    
\citet{aarle11} classified it as a semi-regular variable based on MACHO and OGLE-II data, but did not list a period.
The fact that \citet{fraser08} find the same two periods based on a different data set (MACHO) increases the confidence in the period results in spite of the inferior fit to the light curve.
While the object does not change much in color, $\Delta${\it (V$-$I)} = 0.1, it shows a clear trend of being redder when fainter (Figure~\ref{color}).

\placefigure{obj10_LC}

(13) {\it J052520.76-705007.5} (Figure~\ref{fig2b}):
This object is bright and red, with $<${\it I}$>$ =14.19 
and $<${\it V$-$I}$>$ =1.21.  There is a slight general trend of decreasing brightness over the eight seasons of about 0.02 mag ({\it I}).  The entire peak-to-peak range in the brightness is 0.13 mag ({\it I}), with a typical range in a season of 0.08 mag ({\it I}).  Examination of the individual seasonal light curves suggests some cyclical variability with a timescale ranging from $\sim$10$-$60 days and typical values $\sim$20$-$40 days.  A period analysis of the observed data results in a period of 34.5 days, with secondary values of 169 and 43 days; similar values are determined if we first remove the linear trend.  If we instead seasonally-normalize the  brightnesses, and then search for periodicity, we again determine a value of 34.5 days, with secondary values of 43 and 53 days and the period of 169 days absent.  However, the first of these period values falls slightly below the adopted level for significance and the others fall well below that; also, when taken together,  they do not give a good fit to the complex light curve variability. This object is listed as LMC$-$LPV$-$53645 in the OGLE-III Catalog of Variable Stars, where it is classified as a Long Period Variable or an OGLE Small Amplitude Red Giant (LPV-OSARG).  The periods listed are 34.5, 169, and 43 days, the same as we found.  
An analysis of the MACHO data reveals significant periods of 42.8, 44.5, and 47.2 days; these are in agreement with the values  of 42.8 and 44.3 days found by \cite{fraser08}.  These similar to the secondary period that we find in the OGLE-III data.
\citet{spano11} claim no periodicity in their EROS-2 data and \citet{aarle11} classified it as an object $``$without evidence of any significant light variation.$``$
We regard as uncertain the period value of 34.5 days.
Rather, we suspect that there is not a stable dominant period, but that the object varies with a periodicity between 34 and 45 days.  
The object shows a slight trend of being redder when fainter (Figure~\ref{color}), but this is based on only 14 points and a small range in color ($\Delta${\it (V$-$I)}=0.045 mag).  

(16) {\it J053250.69-713925.8} (Figure~\ref{fig2b}):
The object is very bright, with $<${\it I}$>$ = 13.82,  
and is red, with $<${\it V$-$I}$>$ = 1.25. 
The $<${\it I}$>$ light curve appears to show a small range of variations in the mean light level from season to season over a range of 0.02 mag and a range of a factor of two in the seasonal amplitudes, from 0.06 mag in 2001 to 0.14 mag in 2008.
An examination of the light curve shows a clear cyclical variation in every season but year 5 and also suggestions of a beat period.
Analyzing the data as observed and then normalized yielded the same periods, with a strong period of P$_1$ = 90.7 $\pm$0.1 days and secondary periods of 85.9, 80.8, and 117.8 days.  In the non-normalized data there was also a period of 316 days due to the slight change in the annual mean values.  These results are shown in Table~\ref{periods}.
The power spectrum and phase diagram for P$_1$ based on the normalized light curve are shown in Figure~\ref{Pspec}, 
and the fit to the four periods are shown in Figure~\ref{obj22_LC}.
It can be seen these periods and amplitudes fit the normalized light curve reasonably well.
These results agree with the analysis in the OGLE-III Variable Star Catalog, where it is listed as long-period, semi-regular variable with periods of 90.8, 85.8, and 80.7 days.
\citet{spano11} claimed three periods of 93.3, 412, and 215 days; the first of these is close to the dominant period in the OGLE-III data, as are the two periods of 92.3 and 91.7 days found by \citet{fraser08}. 
\citet{aarle11} classified it as a semi-regular variable, based on MACHO and OGLE-II light curves but did not list a period.
There is a clear correlation with color, with the object redder when fainter (Figure~\ref{color}).

\placefigure{obj22_LC}

(19) {\it J054054.31-693318.5} (Figure~\ref{fig2b}):
The object is faint, with $<${\it I}$>$ = 18.3 mag, and increases systematically in brightness by 0.8 mag over the eight years of observations.  Most of this increase occurs during the first season, during which it increased by almost 0.5 mag and then decreased by 0.1 mag at the end of the season.  During the second season, it increased by 0.16 mag, with a total increase of 0.6 mag in the first two seasons.  Thereafter it increased approximately linearly by 0.13 mag over the next six seasons.  The first four observations of season 3 (2003) appear to be anomalous, as they are $\sim$0.2 mag fainter that the rest of the season.  There are no OGLE ${\it V}$ measurements.  Published photometry indicates that the object is red, with ${\it V}$ = 21.76 mag \citep{fraser08} and thus  $<${\it V$-$I}$>$ = 3.47 mag.  
Possible causes for this rapid increase in brightness during season one are explored briefly in Section~\ref{discussion}.

If we disregard for the period analysis the data of seasons one and two and the first four observations of season 3, then the ${\it I}$ light curve shows a season-to-season range in amplitudes of 0.16 to 0.29 mag, with an average value of 0.24 mag.
Visual inspection of seasons three to eight shows a cyclical variation in the light curve, but with cycle lengths ranging from $\sim$100 to $\sim$150 days.
We carried out a period analysis of the last six seasons, neglecting the first four anomalous observations of season 3.  The data were normalized to the seasonal means to remove the brightening trend.  This resulted in P$_1$ = 144.6 $\pm$ 0.6 days, with additional periods of 130.7 and 168.1 days.
These results are listed in Table~\ref{periods}.
The frequency spectrum, showing the three significant periods, along with the light curve phased to P$_1$ are  displayed in Figure~\ref{Pspec}.
In Figure~\ref{obj22n_LC} is shown the observed light curve for seasons 3 to 8 fitted with the three periods.  The fit is reasonable, given the complicated nature of the light curve.
This object is not included in the OGLE-III Variable Star Catalog.  However, it is listed by \citet{fraser08} as having much different period values, P$_1$ = 5.16 and P$_2$ = 11.16 days based on MACHO data, 
while \citet{spano11} found no period in their EROS data.

\placefigure{obj22n_LC}

\section{NEW SPECTRA}
\label{newspectra}

Optical spectra of three of the LMC objects were obtained at Gemini South on 2009 Dec 5 (no. 12) and 2010 Feb 6 (nos. 8 and 9) using the GMOS instrument under the poor weather program GS-2009B-Q-97. All the observations were taken using the R400 grating and central wavelengths of 6000 and 6050 \AA~ to allow interpolation over the chip gaps. The slit width was 0.75\arcsec. The observations were done in conditions of fair to poor seeing but clear, or nearly clear, skies. 
Details of the reduction procedure and discussion of the individual spectra can be found in Appendix II.

Figure~\ref{spectra} shows the extracted spectra for the three PPN candidates and for two bright, nearby LMC field stars. The spectra are not corrected for the radial velocities of the objects. 
The spectral resolution was 3.4 \AA, well suited to spectral classification.  

\placefigure{spectra}

As expected from the photometry, the optical spectra of object no. 12 and especially no. 8 show heavily reddened continua. The low signal levels in the blue near 4000 \AA~ make classification of the spectra difficult since most of the primary classification indices are at these short wavelengths.

Object 12 is classified as type F8 I(e) based upon the G-band/H$\gamma$ ratio, the H$\alpha$/H$\beta$ line strengths, and the line widths.  
Object 8 has a spectrum that is even weaker in the blue and consequently we are not able to reliably classify it.  The star is classified as type F3 II(e) by \citet{aarle11}, and our spectrum is consistent with this; this spectral type is listed in Figure~\ref{spectra}.
Object 9 shows strong H$\alpha$ emission along with H$\beta$ and the [N II] $\lambda\lambda$6548/6584 \AA~ nebular lines on a relatively blue continuum. 
Based on the weak absorption lines of He I, He II, O II, N III, and C III and the narrowness of these absorption lines, we classify the spectrum as B0.5 I[e].

\section{DISCUSSION}
\label{discussion}

\subsection{Period Studies}

All 22 of these sources are found to be variable, with total ranges (peak-to-peak) over the observing intervals of 0.1$-$1.0 mag in {\it I} and 0.1$-$0.4 mag in {\rm R}$^\prime_C$, based on the available OGLE (19 objects) or MACHO (3 objects) data. 
Some show large ($\ge$0.1 mag) monotonic increases or decreases in brightness, while others show systematic increases and decreases over the observing interval.  Finally, one (no. 20) shows a large increase (0.5 mag) in a single season.
In addition to such long-term changes, they each show shorter timescale variations that we attribute to pulsations, with peak-to-peak (seasonal) variations that range from 0.09 to 1.0 mag in {\it I} (see Table~\ref{var}).

\placetable{var}

Eight of the sources showed periodic light variations.  These, with their dominant periods, are nos. 1 (142 days), 2 (96 days), 4 (157 days), 5 (49 days), 8 (112 or 96 days), 12 (74 days),  16 (91 days),  and 19 (145 days). 
This is similar to the range of periods of 38 to 153 days found by \citet{hri10} for Galactic C-rich PPNs of spectral types G8 to F3.
In addition, object 13 shows a possible periodicity (34.5 days), but is not quite at the level to be considered significant.  
Of the other 13 objects, 
four (nos. 3, 7, 11, and 14) display variations on the order of 10$-$35 days,
two others (nos. 21 and 22) display variations on the order 10 days or less, 
and seven others display variations on a shorter timescale of a few days or less (see Table~\ref{var}).
As noted earlier, most of these, six of eight, had previously published periods that were in agreement with what we found based (usually) on larger data sets, while the other two were very different.

All of those with dominant periods also have additional, rather similar periods that beat against each other and result in the changing amplitudes seen from season to season.  
The ratios of P$_2$/P$_1$ for these objects are recorded in Table~\ref{periods}, where in some cases we have listed the two values obtained using different filters or different time intervals.   
The ratios range from 0.81 to 1.16, with an average value of 0.97.
In the case of object 2, the values of P$_2$ and P$_3$ are reversed, depending upon whether one uses the combined OGLE-II and III data or only the OGLE-III data.  
In the case of object 8, the values of  P$_1$ and P$_2$ are reversed between the R$_C$ and {\it V} light curves.  
Thus there are some cases in which the exact value of P$_2$/P$_1$ depends on the data set used.
If we determine the average values for those in which P$_1$ $>$ P$_2$, we find that P$_2$/P$_1$ = 0.89.
If, on the other hand, we  determine the average values for those in which P$_2$ $>$ P$_1$, we find that P$_1$/P$_2$ = 0.92.  
These ratios are similar to the values of 0.86$-$1.06 found in several Galactic C-rich and O-rich PPNs  and other post-AGB objects \citep{ark10,ark11,hri13,hri15,kiss07,vanwin09}.  
These close period values presumably result from the different pulsation modes excited in these large, intermediate-mass stars.
Note that these are different than the ratios of 0.71$-$0.73 found for double-mode, more massive classical cepheids in the MWG and LMC/SMC \citep{bea97}.

The multi-period fits to the light curves appear to vary from good (objects 2 and 4) to poor (object 5).  However, a part of this has to do with the range of the variation.  There is a clear correlation between the amplitude of the variation and the average residual of the fit.  The fits for objects 8, 11, and 12 do not appear to be very good, but they actually have the smallest average residuals.  
And one can see in many cases that the observed light curve fits tend to produce good timings for the light curve maxima and minima but the amplitudes are often not well fit.  One could expect a better fit if the amplitudes were permitted to change with time.  The complex nature of PPN light curves is not surprising, given that shocks are produced as they pulsate.  These are predicted based on modeling of the pulsation \citep{fok01} and has been observed spectroscopically \citep{leb96,zacs09}.  

\subsection{Correlations of the Period with Temperature and Brightness Variability}

\citet{hri10} found a clear correlation for Galactic C-rich PPNs between pulsation period and both the effective temperature and the amplitude of the variation.  
Such correlations are expected since PPNs evolve with time toward higher temperatures as the stars decrease in size due to both the fusion of hydrogen at the base of the atmosphere and the presence of a fast wind removing the outer envelope.  
Thus one expects the surface gravity to increase as the star evolves from G to B spectral type, and assuming that the pulsations are radial, this will cause the period to decrease with time.

We have examined the objects in our present sample that have periods to see if such relationships exit for LMC/SMC objects.  This requires that we have T$_{\rm eff}$ for those with periods.
A detailed, high-resolution abundance study of object 2 has been carried out by \citet{smedt12}.  They found the object to indeed be C-rich, with C/O = 1.9$\pm$0.7, iron poor ([Fe/H]=$-$1.34$\pm$0.32), and with a high abundance of s-process elements.  From their atmospheric model, they determined T$_{\rm eff}$ = 6250 $\pm$ 250 K.  
Similarly, \citet{aarle13} carried out a detailed high-resolution abundance study of objects 5, 12, and 16 and determined T$_{\rm eff}$ values of 5750, 6750, and 5500 K, respectively. 
They found them each to be C-rich, with C/O abundance of 2.1 (average, based on two different spectra observed one month apart), 1.5, and 2.5, respectively, and all iron poor ([Fe/H] = $-$1.2) but with a high abundance of s-process elements. 
These are common properties of Galactic PPNs \citep{vanwin00}.
Thus four of the eight PPNs with well-determined periods have temperature determinations based on high-resolution spectra and model atmosphere analyses (nos. 2, 5, 12, and 16).  
Additional spectra of three objects at lower spectral resolution (1500$-$3000) have been obtained and analyzed by \citet{haj15} using model atmospheres.  For two of these we have been found periods.
All three were determined to have low gravities (log~{\it g} = $-$0.5) and low metallicities ([M/H] $\approx$ $-$1.0).  
They determined temperatures for objects 8 and 12 as 6750 and 6250 K, respectively. 
For object 12, for which there are two temperature determinations, we will use the value of 5750 K based on higher-resolution spectra in our discussion of possible correlations, but it may simply be that the higher value of 6250 K was obtained when the object was brighter and bluer.
Object 1 has been classified as a carbon star or post-AGB star on the basis of its infrared excess and its optical spectrum \citep{white89,groen98}.  However, a comparison of its spectrum with those of carbon-rich galactic PPNs
suggests that it shows the weaker C$_2$ absorption features commonly seen in these PPNs, 
although in contrast it possesses an unusually strong H$\alpha$ emission feature.  
On the basis of its similarity to IRAS Z02229+6208 \citep{hri99}, we classify its spectrum as approximately G8$-$K0 0-Ia 
and assign it an approximate temperature of 5500 K \citep{red99}.
These temperature values are listed in Table~\ref{results}.

\placetable{results}

The results for the comparison of {\it P} and {\it T$_{\rm eff}$} are shown in Figure~\ref{P-T},
where we have also shown the results of \citet{hri10}\footnote{We have modified slightly Figure 18 of \citet{hri10} to include the temperature of 5000 K determined by \citet{kloch06} for IRAS 20000+3239 , rather than an estimate based on its spectral type.} for 12 C-rich PPNs in the MWG.  
Note that they found a linear relationship over the temperature range 5000$-$8000 K, and empirically determined a rate of decrease of $\Delta$$\it P$/$\Delta$$\it T$$_{\rm eff}$ = $-$0.047 days K$^{-1}$.
The sample size is much smaller in the present study; there are only four LMC and two SMC objects and the range in temperature is more limited than in the MWG. 
These four LMC PPNs also show a trend of decreasing {\it P} with increasing {\it T$_{\rm eff}$}.
Comparing the two samples, one sees that three of the LMC values lie to the left of those of the MWG objects, at systematically lower temperatures or shorter periods. 
This sample is small, so the differences may not be significant. 
If it is real, however, one possible explanation that comes immediately to mind is the difference in metallicity between the two galaxies, with the metallicity of the LMC about 0.4 of the solar metallicity \citep{cox00}.  But PPNs are known to be metal poor compared to the Sun, \citep{vanwin00,aarle13}, 
so one should compare directly the metal abundance for the two samples.
For the three LMC objects analyzed by \citet{aarle13}, the metallicities are [Fe/H] = $-$1.2 with very little variation.  This can be compared with the values of those in the MWG.  Similar high-resolution spectral analyses of the MWG sample show a range of [Fe/H] = $-$0.3 to $-$1.0, with an average value of $-$0.7 \citep{vanwin00,red99,red02}.  
This difference in [Fe/H] of $-$0.5 implies a metallicity of these C-rich LMC PPNs of one third of the value of those in the MWG.  
Thus an apparent offset could occur if, at the same {\it T$_{\rm eff}$}, a lower metallicity resulted in a shorter period of pulsation. 

\placefigure{P-T}

The relationship of pulsation period with amplitude can also be examined.  \citet{hri10} compared the maximum seasonal peak-to-peak $\Delta${\it V} amplitude with {\it P}, and found a general, direct, although not linear relationship between $\Delta${\it V} and {\it P}; the objects with {\it P} greater than 120$-$130 days possessed significantly larger amplitudes (generally 0.5$-$0.6 mag), while those with shorter periods had smaller amplitudes (0.13$-$0.23 mag). 
We have plotted in Figure~\ref{P-Amp} these parameters for the eight LMC/SMC objects that have periods, where we have also reproduced the relationship found for the MWG \citep[Fig. 19]{hri10}. 
For this present sample, most of the individual points are in agreement: the object with the longest period has the largest amplitude and those with shorter periods have smaller amplitudes in the range 0.14$-$0.23 mag.
However, there are two objects in this present sample that are not in agreement with this trend.
The SMC object 1 at P = 96 days and $\Delta${\it V} = 0.40 mag has an amplitude that is about twice as large as  other objects with periods of $\sim$100 days.
LMC object 1, with a large period of 142 days, has a small $\Delta${\it V} amplitude of only 0.15 mag.  This is anomalous since it is even smaller than the $\Delta${\it I} amplitude of 0.21 mag.  In all of the other objects in this sample and in the study by \citet{hri10}, it was found that the amplitude was larger at shorter wavelengths. 
We propose that this deviant value is due to the fact that there are many fewer V than I observations, by about a factor of ten, which in this case has conspired to not show the full amplitude in V.  
We attempted to empirically correct this by scaling the $\Delta${\it I} amplitude by the typical value of the 
ratio of the amplitude range in {\it V} to that in {\it I} for the other four with OGLE {\it V} and {\it I} data.
This ratio has an average value of 1.5, and then multiplying the $\Delta${\it I} value by this one gets the scaled $\Delta${\it V} value. 
We do the same with object 22, which has no OGLE {\it V} data.  These are plotted in Figure~\ref{P-Amp}, with distinguishing symbols.

\placefigure{P-Amp}

Examination of Table~\ref{var} does raise the question about the large amplitudes measured for some of the non-periodic variables.
Three of the four objects with the largest seasonal peak-to-peak variations are not long-period variables.  
Rather, no periods have been found for them, which may seem inconsistent with the trend discussed above.  
However, these three are among the faintest sources in this study ($<$${\it I}$$>$ $\ge$ 19.0 mag) and they possess the largest photometric uncertainties.  This level of faintness and photometric uncertainty is the likely cause of the large ranges in their brightness.  The fourth of these with the large seasonal value of $\Delta$${\it I}$ is the periodic object with the longest period (157 days, no. 4), consistent with the trend discussed above.

\subsection{Pulsational Temperature Changes}

Most of the objects (14 of 19) with multicolor observations are found to be redder when fainter 
(see Fig.~\ref{color}).  This is indicated these in Table~\ref{var}.
This relationship has previously been found for all of the C-rich F$-$G PPNs in the MWG \citep{hri10}.  This is attributed in most cases to changes in temperature as the stars pulsate.
We have investigated this further for those that have temperature determinations based on high-resolution spectra and model atmosphere analyses.
The observed color variations were used to find the range in temperature variations.  Comparison of the times of the spectroscopic observations with light curves revealed the pulsation phases, from which we determined the actual temperature changes.
Objects 12 and 16 had good light curve coverage at the times of the spectra, and were at approximately minimum and slightly above mid-level of the light curves, respectively.  For object 2, the light curve coverage stops about three-fourths of a cycle before the spectroscopic observation, but the data are so well fitted by the model curve that we feel confident that the spectroscopic observation was made when the star was at mid-level in its light curve.  For object 5, the only other object in the sample with a temperature determined from high-resolution spectra, the photometric observations derive from the MACHO data set, which ended some eight years prior to the spectroscopic observation.  We have not tried to extrapolate to find the phase at the time of the spectrum.
We used the color$-$temperature relations for supergiants as listed by \citet[][Table 15.7]{cox00}.  Since this uses ({\it V$-$I}) colors on the Johnson system, we converted our observations from the Cousins system using the relationships given by \citet{fernie83}.
These temperature ranges are listed in Table~\ref{Tvar}; they vary from 350 to 1100 K and represent the maximum temperature changes in a season for these three stars. 
In a detailed photometric and radial velocity study for two MWG C-rich PPNs, \citet{hri13} found  that they were  largest in size when faintest and coolest, and that they were smallest when brightest and hottest.  While it is likely that these LMC/SMC C-rich PPNs pulsate in a similar fashion, it would be useful to confirm this for a few of the brightest ones with contemporaneous radial velocity and light measurements.
These three LMC/SMC objects have rather modest changes in $\Delta$({\it V$-$I}), ranging from 0.07 to 0.18 mag. 
Five of the objects in the sample have much larger 
color ranges, from 0.5$-$0.9 mag.  
However, these are the five faintest sources (in {\it V}) in the sample, with very large average uncertainties in ({\it V$-$I}) color indices ranging from 0.10 to 0.19 mag, so much of the large color range may be attributed to the large uncertainties in their observations.

\placetable{Tvar}

\subsection{Long-Term Brightness Changes}

Several of the objects show longer-term, often monotonic changes in brightness.  Three (nos. 6, 20, 22) show monotonic increases of $\ge$0.08 mag ({\it I}), 
one (17) shows a large increase (0.19 mag) followed by a large decrease (0.09 mag), while three more (2, 18, 21) show significant changes over several years in one direction, brighter or fainter, over only part of the observing interval. Object 19 shows a a large (0.5 mag) and rapid increase in brightness during season one, followed by a more gradual increase of 0.2 mag over the following seven years.  
Among these eight objects are the four faintest ones (nos. 6, 17, 18, 22), $<$I$>$ $\ge$ 19.0 mag, and for these faint objects some of these trends may reflect large and systematic uncertainties in the data.
However, the other four (2, 19, 20, 21)  are bright to medium-bright, $<$I$>$ = 14.5$-$18.3 mag, with relatively small uncertainties in the individual data points ($\sigma$=0.004$-$0.03 mag).
We have seen such trends in our study of MWG PPNs and suspect that they are due to systematic changes in the line-of-site opacity of the circumstellar dust.  
Decreases in brightness could be due to increases in dust opacity, such as from a recent mass ejection event, and increases would be due to the dissipation of dust through expansion.  
Or either could be due to the passage of a dust cloud across the line-of-site.  

Object 19 is a special case, since it shows a relatively large increase in a single season.  In this it is somewhat reminiscent of the MWG C-rich PPN IRAS 19500$-$1709, although on a more extreme scale.
IRAS 19500$-$1709 showed a sudden decrease of 0.12 mag (V) and increase in reddening ($\Delta$({\it B$-$V)}=+0.05 mag) 
in a single season, followed by a rise of 0.19 mag over the next four years \citep{hri10,ark10}.  
\citet{hri10} attributed this to changing dust opacity.  
Another, more interesting possibility, is that object 19 is coming out of eclipse by a binary companion.  
Unfortunately there are not OGLE-II observations to document the light curve of the previous four years. 
However, to investigate this further, we examined the MACHO data for this object (MACHO 76.9852.1227).  
While these data have a large uncertainty, with  $<$$\sigma$$_ R$$>$ = 0.13
after the removal of some observations with very large uncertainties, they do show a clear pattern over the six (1993$-$1999) years of observations.  The object appears to decrease in brightness by $\sim$0.9 mag over the first four years and then increase by $\sim$0.6 mag ({\it R}) over the next two.  There is then a two-year gap before the OGLE-III data, which then begin with the rapid increase in brightness of 0.5 mag ({\it I}).  
Taken together, these data sets are not consistent with an eclipse.
Rather they are reminiscent of 
the R Coronae Borealis (RCB) phenomenon, with sudden drops in brightness and also low-amplitude cyclical variations.  For RCB stars, which are also C-rich, the drops in brightness are large, up to 8 mag on time scales of a few years, and the low-amplitude variations have typical cycle lengths of 40$-$110 days with amplitudes of a few tenths mag \citep{cla96}. 
The brightness decreases in RCB stars are attributed to dust formation episodes that partially obscure the star.
Object 19 does not have such a large drop in magnitude ($\sim$1), has a longer period of variability (145 day), and has lower amplitude ($\sim$0.05 mag).  So it appears to have some of the same properties, but on a smaller scale, suggesting less dust formed.  

Note that the presence of circumstellar dust is seen not only in the large infrared excesses, the infrared dust features, and the suggested dust-caused monotonic brightness trends, but also in the very red colors of some of these objects.
While we do not have spectral types for most of them to permit us to determine the reddening in the individual objects, we note that ten of them are redder than ({\it V$-$I}) $\ge$ 1.6 mag, with five of these having values larger than 2.0 mag.
A supergiant would have to have a spectral type of K5 or later to be this red due to its photosphere, and
hence there is a reddening of $>$ 1 mag in ({\it V$-$I}) if these red stars are earlier than G5.

\subsection{Correlation of the 21~$\mu$m Feature with Period and Temperature}
 
We also investigated whether there exists a correlation between the presence and strength of the 21 $\mu$m emission feature with either pulsation period or temperature (or spectral type).  We have listed in Table~\ref{results} the presence and strength of the 21 $\mu$m feature, as determined by \citet{volk11}, as either present (Y), weak (W), questionable (Q; which \citet{volk11} classified as very weak or questionable), or not present (N).
For the additional three objects added from the study by \citet{sloan14}, we classified them similarly based on their Spitzer spectra.
The seven with a definite (Y) 21 $\mu$m features consist of three of the six with long periods (70$-$160 days) and Spitzer spectra, three of the five with medium timescale variability (10$-$35 days), and only one of the nine with short timescale variability ($\le$10 days).
The other eight with short timescale variations have weak (1), questionable (2), or no (5) 21 $\mu$m feature.
Thus there appears to be a general correlation between the presence and strength of the 21 $\mu$m feature and the pulsation timescale. 
Those with long pulsation periods or medium timescale variability generally show a definite feature and those with short timescale variability possess no or a questionable 21 $\mu$m feature.
One can compare these results with what is seen in the MWG C-rich PPNs.  Examining the presence and strengths of the 21 $\mu$m feature with their periods for the MWG C-rich PPNs, one sees that all 10 of those with long periods  (70$-$160 days)  have a definite feature \citep[comparing the results of][]{hri08,hri10}.
On the other hand, none of the three hot (early B spectral type) PPNs classified as C-rich based on their Spitzer spectra \citep{cer09}, and which all display short timescale variability \citep{ark04,ark06b,ark13}, possesses a 21 $\mu$m feature.

A correlation also exists with temperature for these LMC/SMC objects, although based on fewer objects.
Four of the five objects in the present study with cooler temperatures (5750$-$6750 K) and F$-$G spectral types have a definite 21 $\mu$m feature, while the two with known early spectral types (nos. 9 and 21) have no 21 $\mu$m feature.
A clear temperature effect has been seen for MWG PPNs.
The 21 $\mu$m feature is almost solely detected in C-rich PPNs of F$-$G spectral types (5000$-$8000 K) and not in hot, early spectral type PPNs (see above) or PNs.  
It appears that the carrier of the 21 $\mu$m emission feature is present in the environment of PPNs of spectral type late-G to late-B, but that as the stars evolve to higher temperatures the carrier is either destroyed or is no longer excited in the same way. 
The correlation of the 21 $\mu$m feature with the timescale of variability is simply the result of the period-temperature relationship discussed earlier.

\subsection{Classifying the Objects}

It seems surprising that so few of these C-rich, post-AGB objects in the LMC/SMC, eight (or perhaps nine) out of 22, show periodic variability, while in the MWG almost all of those with carefully documented multi-year light curves do \citep{hri10}.
In contrast, relatively large number these in the LMC have short-timescale variations.
Such short-timescale variations are also seen in the MWG, but only in hot, early B spectral type PPNs
and early-PNs \citep{ark06}, of which there are only three clearly identified that are C-rich in the MWG.
Only two of these short-timescale LMC variables have a spectral type (nos. 9, 21) and they are classified  hot stars (B0.5~I(e), WC?), in agreement with what is found in the MWG.
We have used the information on periodicity and timescale of variability, temperature and spectral type, 
and the presence or absence of the 21 $\mu$m feature, along with analogy with the PPNs in the MWG, to classify the nature of the stars in the LMC/SMC sample.  

We classify as PPNs the eight with long periods, six of which have model atmosphere and abundances analyses that support this assignment.  Object 19  had once been classified as a young stellar object (YSO), but \citet{sea09}, in a large study of YSOs based largely on their mid-infrared spectra, classified it as a non-YSO, possibly a PPN. While we classify it as a PPN, we note that it does possess ``mild'' RCB-type drops in brightness.
For a PPN, object 1 has an unusually large luminosity, as noted by \citet{sloan14}, but it is not above the limit for an AGB star.
The nine with short timescale variations of $<$10 days we tentatively classify as hot PPNs or young PNs.  Two of these have spectral types that support this (nos. 9 and 21; see Table~\ref{results}).
The five with medium timescale variations, 10$-$35 days, we also classify as PPNs and suspect them to have spectral types of A to late-B.   The two of these that have spectral types support this (nos. 7 and 13; see Table~\ref{var}).
Future spectroscopic observations of the objects in these later two groups will help to confirm or refine these classifications.

Studies of PPNs in the LMC and the MWG suggest a different spectral type number distribution between the two galaxies.
\citet{aarle11} noted that the number of so-called ``shell'' sources in their study of the LMC, which we would classify as PPN candidates, peaks at B spectral type.
However, in the MWG the PPNs, in general, peak in number at the F spectral types \citep{suarez06} and the C-rich ones peak in number in the F-G spectral types \citep{hri08}. 
Thus there appears to be a difference in the spectral type distribution of PPN candidates between the LMC and the MWG, with a much larger fraction in the LMC peaking at higher temperatures.
Our classification results appear to be consistent with this difference in the distribution of PPNs.
Assuming the general correctness of our above classifications, they indicate, at least for C-rich objects, a significantly larger fraction of hot, early (B) spectral type PPNs in the LMC and SMC (9/22) compared to the fraction in the MWG (3/13; based on those with published spectrographic and variability studies).

What could explain this difference in spectral type number distribution between the PPNs in the LMC and MWG?
It could be that those in the LMC evolve more quickly during the early post-AGB phase.  
A higher mass loss rate for those in the LMC would accomplish this, but we have no direct evidence of the rate of mass loss in these objects to support this.  
Systematically higher masses for those in the LMC would also accomplish this, since higher mass objects evolve more quickly \citep{blo95}.  
We have investigated this by comparing their luminosities, since those with higher mass would be expected to have higher luminosities.
\citet{volk11} and \citet{sloan14} have calculated infrared fluxes for the 20 of these objects with Spitzer IRS spectra (thus excluding objects 5 and 16).  \citet{sloan14} calculate the flux in the 5$-$37 $\mu$m region, while \citet{volk11} also include estimations of the contribution at longer wavelengths and in the 2$-$5 $\mu$m region.
Comparisons were then made between six of these 20 objects with well-determined periods and mid-infrared spectra (excluding objects 5, and 16), and both the nine that varied on short ($\le$ 10 days) timescales and the five that varied on medium (10$-$35 days) timescales (see Table~\ref{results}, col. 7).  
It is found that the average fluxes for those six with periods are significantly lower than the averages for the short or medium timescale groups.  The flux ratios were 0.46 (periodic/short), 0.71 (periodic/medium), and 0.54 (periodic/(short+medium)).  This flux difference suggests that those with the shorter timescale variability indeed have higher masses, which has led to their faster evolution to their early B spectral type range.
However, their faster evolution to the B spectral types must be accompanied by a means to slow down the evolution in the B spectral type range.  
While the cause for such a later slowdown in evolution is not known, nevertheless, the fact remains that distribution of PPNs in the LMC appears to peak at higher temperatures than it does in the MWG.
It has also been found in general that the dust shell temperatures of the C-rich PPN candidates in the LMC are significantly higher than those in the MWG \citep{volk11}. This is consistent with the idea previously suggested by \citet{volk11}, that the LMC stars evolved more quickly and that the dust shell is correspondingly closer to the star for a given spectral type.  However, if the LMC PPNs are more luminous, on average, then that could also contribute to the higher dust temperature.

The inclusion of the above discussion is motivated by the tentative assignment of the short timescale LMC/SMC variable objects as hot, early B spectral type PPNs and early PNs.  
Spectroscopic observations to test this assignment are desirous, but will not be easy to obtain.  Only two of the short timescale variables without spectral classifications are brighter than {\it V} = 20 mag; objects 7 and 16 have {\it V} $\approx$ 16 mag.

\section{SUMMARY AND CONCLUSIONS}

In this paper, we have analyzed the photometric variability of a sample of 22 C-rich PPNs from the LMC and SMC.  We then compared them with a sample consisting of the majority of the know C-rich PPNs in the MWG. 
The primary results are listed below.

1. All 22 of these C-rich post-AGB objects in the LMC/SMC are found to vary in brightness.  Some show periodic variation and others show cyclical variation of a medium (10$-$35 days) or short (less than few days) timescale.  Some also show long-term trends in brightness over several years, although these are generally small compared to the maximum seasonal variations.

2. Periods were determined for eight of the objects, ranging from 157 to 49 days.  These agree with the period range found for a sample of 12 C-rich PPNs in the MWG, which ranged from 153 to  38 days \citep{hri10}.

3. The eight periodic objects are each found to have multiple periods, ranging from two to four.  This was expected from the light curves, which show strong modulation in the pulsation amplitudes suggestive of beat periods due to closely spaced periods.  The secondary periods ({\it P$_2$}) determined are indeed close to the primary periods ({\it P$_1$}), sometimes a little shorter and sometimes a little longer, with an average ratio of secondary to dominant of {\it P$_2$}/{\it P$_1$} = 0.97.

4. Correlations are found between the dominant pulsation period ({\it P$_1$}) and effective temperature ({\it T$_{\rm eff}$}), with {\it P$_1$} decreasing with increasing {\it T$_{\rm eff}$}, and between {\it P$_1$} and the pulsation amplitude, with amplitude decreasing with decreasing {\it P$_1$}.  Again, these are in agreement with what is found for the sample of MWG objects, and they are what one would expect based on the evolution of stars in this PPN phase.  However, for the {\it P$_1$}$-${\it T$_{\rm eff}$} relationship, there is a suggestion that the LMC objects are shifted systematically to lower {\it T$_{\rm eff}$} and/or shorter {\it P$_1$}.  However, this tentative suggestion is based on only four objects.

5. A correlation appears to exist between the presence of the 21 $\mu$m spectral feature 
feature and both the pulsation timescale and the temperature (or spectral type), although this is based on a small number of objects.  Those with long periods generally have a 21 $\mu$m feature, as do the majority of those with medium timescale variability (10$-$35 days), while those with short timescale variations do not.  Likewise, those with cooler temperatures and later spectral types tend to have the 21 $\mu$m feature and the hotter objects do not.  These are in agreement with the correlations found in the MWG PPNs,
and that are attributed to the temperature environment necessary to produce or maintain the carrier of the feature and to also excite it.

6. The 14 without determined periods vary on a medium (5; 10$-$35 days) or short (9; $\le$ 10 days) timescale.  The short timescale is suggestive of hot PPNs (early B spectral types) or young PNs, since that is what has been found in the MWG.  
In the MWG, the majority of C-rich PPNs have spectral types of F or G.
In contrast, based on the above classification of the short timescale pulsators as hot PPNs, the majority in the LMC appear to be hot PPNs.
The cause of this difference is uncertain, but it is suggested that they may evolve more quickly due to higher masses or higher mass loss rates.
It would be useful to obtain spectroscopic observations of the brighter of these without determined periods to confirm the suggested spectral classifications as early-B spectral types.  The two short timescale variables that do have spectral classifications are indeed both hot objects. 

7. One object (J054054.31-693318.5) appears to show some RCB characteristics, but on a smaller scale.

While the main period results are firm, the correlations with other physical parameters are less so.
They would profit by the increase in the number of PPNs with periods, in order to seek confirmation of the trend and of the suggested offset in {\it P}$-${\it T$_{\rm eff}$}.
And they would also profit by the determination of spectral types for more of the objects, so that one can investigate with a larger sample the classification of the short timescale variables as hot, B spectral type objects.
The discussion of the apparent differences in the spectral type distribution of PPNs in the LMC and the MWG was predicated on this spectral classification of the short timescale variables.  
However, at present there are not good prospects to significantly enlarge this sample.  
We are lacking a facility such as {\it Spitzer} that can obtain mid-infrared spectra to help identify such C-rich candidates.  
Thus we must resort to techniques such as visible-band spectral searches of color-selected mid-infrared excess objects.  A first step in this would be to obtain spectral types for the five objects in the present sample that lack temperature classifications that are brighter than {\it V} = 19 mag.  This sample consists for two that have short timescale variations and three with medium timescale variations, with two in the later group showing definite evidence of the 21 $\mu$m feature.

\acknowledgments  

We thank K. De Smedt for sending us the time of the observation of J004441.
The comments of the referee have helped us to sharpen the presentation and 
better support the results, and are gratefully acknowledged.
BJH acknowledges support from the National Science Foundation
(AST 1009974, 1413660). 
RS and MH acknowledge support from NCN grants 2011/01/B/ST9/02031 and
2011/01/D/ST9/05966.
We acknowledge support from the EU FP7-PEOPLE-2010-IRSES program in the
framework of project POSTAGBinGALAXIES (Grant Agreement No.269193).
The OGLE project has received funding from the European Research Council under the EU FP7/2007-2013 program (Grant Agreement No.246678).
Some of these observations were obtained at the Gemini Observatory, which is operated by the
Association of Universities for Research in Astronomy, Inc., under a cooperative agreement
with the NSF on behalf of the Gemini partnership: the National Science Foundation (United
States), the National Research Council (Canada), CONICYT (Chile), the Australian Research 
Council (Australia), Minist\'{e}rio da Ci\^{e}ncia, Tecnologia e Inova\c{c}\~{a}o (Brazil) and
Ministerio de Ciencia, Tecnolog\'{i}a e Innovaci\'{o}n Productiva (Argentina).
This research has made use of the SIMBAD database, operated at CDS, Strasbourg,
France, and NASA's Astrophysical Data System.

\clearpage

\section{APPENDIX I: PHOTOMETRIC VARIABILITY OF THE NON$-$PERIODIC  OBJECTS}
\label{NonP}

\begin{figure}\figurenum{2a}\epsscale{1.10} 
\plotone{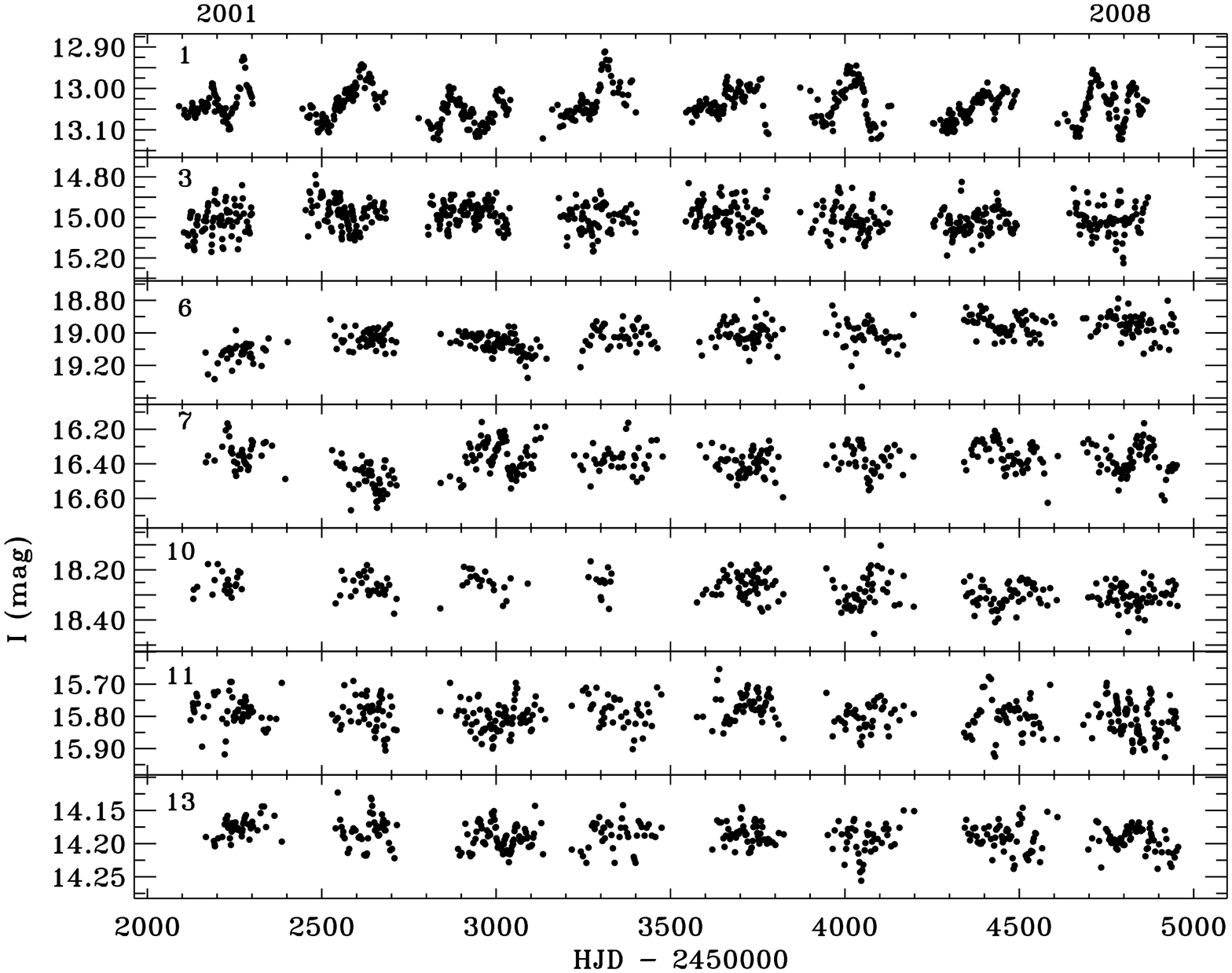}
\caption{The {\it I} light curves for seven objects with OGLE-III data, from 2001 to 2008.  The numbers in the upper left of each panel refer to the object identifications listed in Table~\ref{object_list}.
\label{fig2a}}
\epsscale{1.0}
\end{figure}

(3) {\it J010546.40-714705.27} (Figure~\ref{fig2a}):
This object is bright and relatively blue, with $<${\it I}$>$ = 15.00 and $<${\it V$-$I}$>$ = 0.39.  Its mean {\it I} light level is approximately constant from season to season over the 8 years of observations.  The overall light curve varies within a range of ~0.43 mag (peak-to-peak), with seasonal variations lying between 0.22 and 0.36 mag.  The variations appear to show a generally cyclical behavior in a season, with cycle lengths varying from $\sim$ 10-30 days, but no periodicity was found in the data.  There is a small range of variation in color with no apparent correlation with brightness (Figure~\ref{color}). 

(6) {\it J050713.90-674846.7} (Figure~\ref{fig2a}): 
This object is faint and red, with $<${\it I}$>$ = 19.02 and $<${\it V$-$I}$>$ = 2.48.  It shows a general increase in seasonal brightness of $\sim$0.2 mag ({\it I}) over the 8 seasons of data.  Within a season, it varies by ~0.20$-$0.25 mag ({\it I}).  An examination of the data season by season shows short timescale variability on the order of a few days or less with no apparent cyclical behavior, and the period analysis shows no periodicity in the data.  
It shows a very large variation in color, $\Delta$({\it V$-$I}) = 0.9, and a clear trend of being redder when fainter (Figure~\ref{color}).

(7) {\it J050835.91-711730.6} (Figure~\ref{fig2a}): 
The light curve shows a large range of variation of 0.50 peak-to-peak ({\it I}), with a typical variation of 0.35 mag in a season.  The light curve is brighter during the first season, decreases by $\sim$0.15 mag in the second season, increases in the third season back to the level of the first season, and then is a little fainter ($\sim$0.03 mag) during the next five seasons.  An examination of the light curve season by season shows some suggestions of a cyclical variation 
with short timescales of 8$-$20 days and some longer ones of 20$-$60 days, but with lots of scatter and no regular pattern.  The light curve data were analyzed, both as observed and with the data normalized season by season, but no significant periods were found.  
\citet{fraser08} claim periods of 363 (one-year artifact?) and 71.6 days in the MACHO data and 
\citet{spano11} claim periods of approximately 496, 133, 195, and 88.7 days in the EROS-2 data.
The strongest periodicity that we find is 209 days, but that falls slightly below our significance limit.
We agree with \citet{sos09,sos11} in claiming no periodicity in the data.
This object is relatively bright, with $<${\it I}$>$ = 16.38 and $<${\it V$-$I}$>$ = 1.96.  There is a general trend of the object being redder when fainter (Figure~\ref{color}).  
The spectral types listed for this object in Table~\ref{var} cover a large range, late-B to early-G \citep{mat14}.
This spectral range would lead to an expected color range of $\Delta$({\it V$-$I}) = 0.96 mag on the Johnson system, which converts to a range of $\sim$0.7 on the Cousins system.  This is much larger than the full range of 0.12 mag observed.  While the ({\it V$-$I}) data set is small, 13 points spread over six seasons, still this large a range in spectral type appears to be  inconsistent with the color variation observed.

(9) {\it J051228.18-690755.8} (Figure~\ref{fig1}):
This object is bright and relatively blue, with $<${\it I}$>$ = 16.13 and $<${\it V$-$I}$>$ = 0.31.  It has been observed in both the OGLE-II and OGLE-III surveys. The object shows variability within a season over a typical range of 0.31 mag ({\it I}), with a peak-to-peak variation over the entire 13 years of observation of $\sim$0.5 mag.  The variations within a season are of a short timescale, a few days or less, with no cyclical variation seen by visual inspection.  The formal period analysis does not determine a periodicity in the data.  The color variation is small, and does not appear to be correlated with brightness (Figure~\ref{color}).

(10) {\it J051828.18-680404.0} (Figure~\ref{fig2a}):
It shows an average variation of 0.20 mag ({\it I}) over a season and average values of $<${\it I}$>$ = 18.28 mag and $<${\it V$-$I}$>$ = 2.16. The object varies in light over a short timescale of several days or less, but no period is revealed in the data.  The seasonal average values of the {\it I} light curve are approximately constant for the first five seasons but then decline from 2006 to 2008 by $\sim$0.04 mag ({\it I}).  There are ({\it V$-$I}) color data from the last four seasons and they show a large variation in color, $\Delta$({\it V$-$I}) = 0.5, being redder when fainter (Figure~\ref{color}).

(11)  {\it J051845.23-700534.5} (Figure~\ref{fig2a}):
 This object is relatively bright and red, with $<${\it I}$>$ = 15.80 and $<${\it V$-$I}$>$ = 1.91.  The light curve shows a peak-to-peak range of 0.27 mag, typically varying over 0.21 mag in a season with no long-term, monotonic trends in the light curve. 
A close examination of the individual seasons shows the appearance of cyclical variations of $\sim$10-35 days, most prominently seen in 2008. 
A period analysis of the observed data yields a period of $\sim$480 days, but this goes away if we normalize the data to the seasonal average light levels.  Additional periods of 25.7 and 35.1 days are found, but none of these three are formally significant.   
It may be that there are multiple or changing periods in the range of 10 to 35 days in the data, but there are no dominant periods over the entire eight years of observations.  These period values are similar to those listed in the OGLE-III Catalog of Variable Stars, where it is listed as LMC$-$LPV$-$39671 with periods of 38.7, 480, and 25.7, and classified as a long-period variable or a semi-regular variable.  While periods like these are formally determined by the frequency analysis, they are not significant and they do not give a good fit to the light curve.  
\citet{spano11} claim five periods in their EROS-2 data, 41.1, 165, 35.1, 59.3, and 320 days, while 
\citet{fraser08} find no periodicity in the MACHO data.
The color does not appear to be correlated with brightness (Figure~\ref{color}).  

(14) {\it J052902.39-661527.7} (Figure~\ref{fig3}): 
This object was not observed in the OGLE surveys, but it was observed in the MACHO survey.  It is of medium brightness, $<${\it I}$>$ = 17.35 and extremely red, $<${\it V$-$I}$>$ = 4.0  \citep{zar97}.  MACHO data are available in the {\it R} filter only.  The light curve data vary over a range of 0.35 mag, following the removal of a few (five) outliers.  While the uncertainly in the data is large, with $<$$\sigma$$>$ = 0.07 mag and $\sigma$$_{max}$ = 0.20 mag, the light curve does appear to vary in brightness.  Visual inspection indicates a timescale of the variation $\sim$10$-$20 days, but this length may be determined partly by the sampling frequence.  The last season visually suggests a period of $\sim$100 day, but this is not seen in the other years.  A formal analysis shows no significant period in the data.

\begin{figure}\figurenum{2b}\epsscale{1.10} 
\plotone{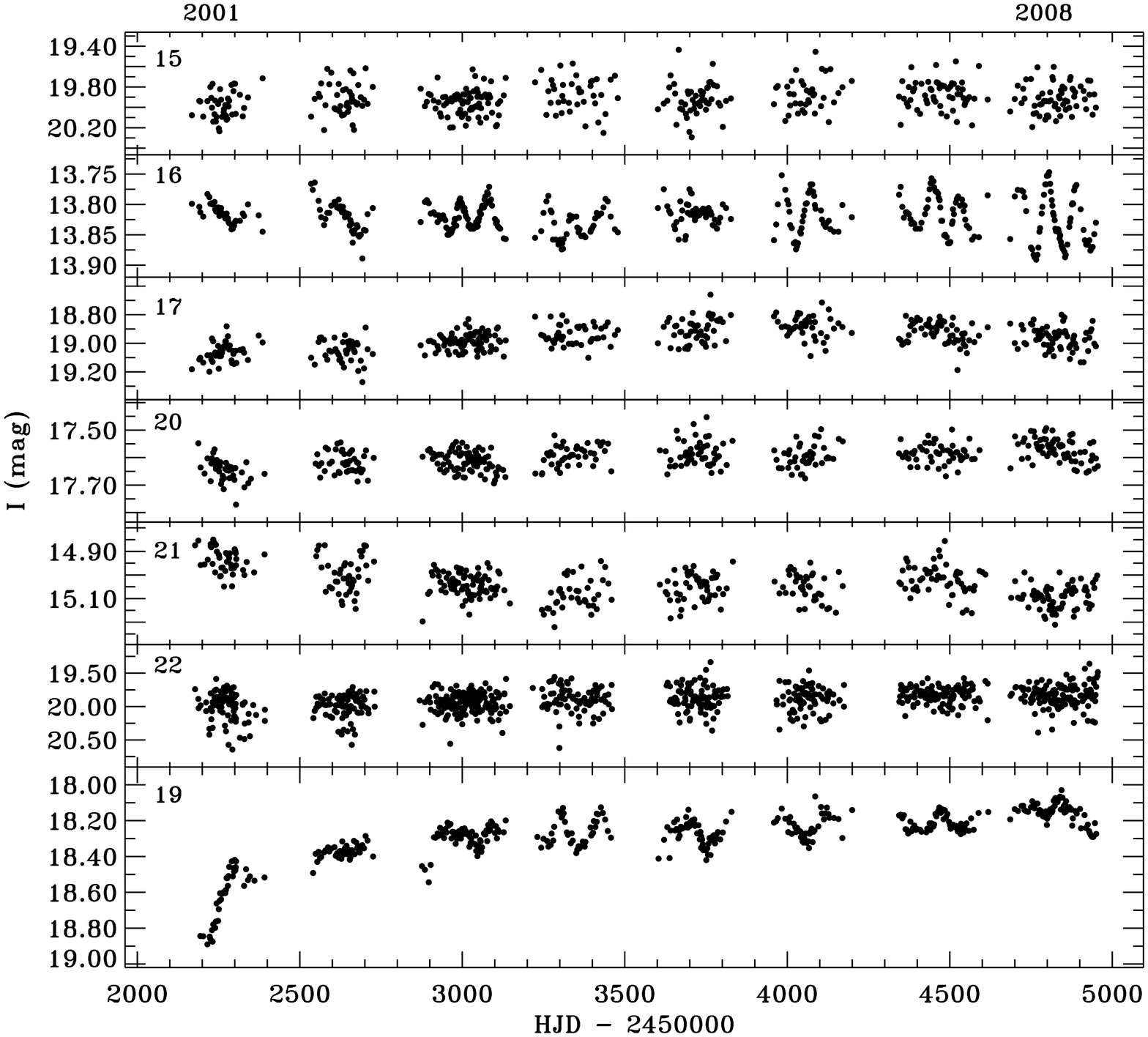}
\caption{The {\it I} light curves for seven objects with OGLE-III data, from 2001 to 2008.  The numbers in the upper left of each panel refer to the object identifications listed in Table~\ref{object_list}.
\label{fig2b}}
\epsscale{1.0}
\end{figure}

\begin{figure}\figurenum{3}\epsscale{1.10} 
\plotone{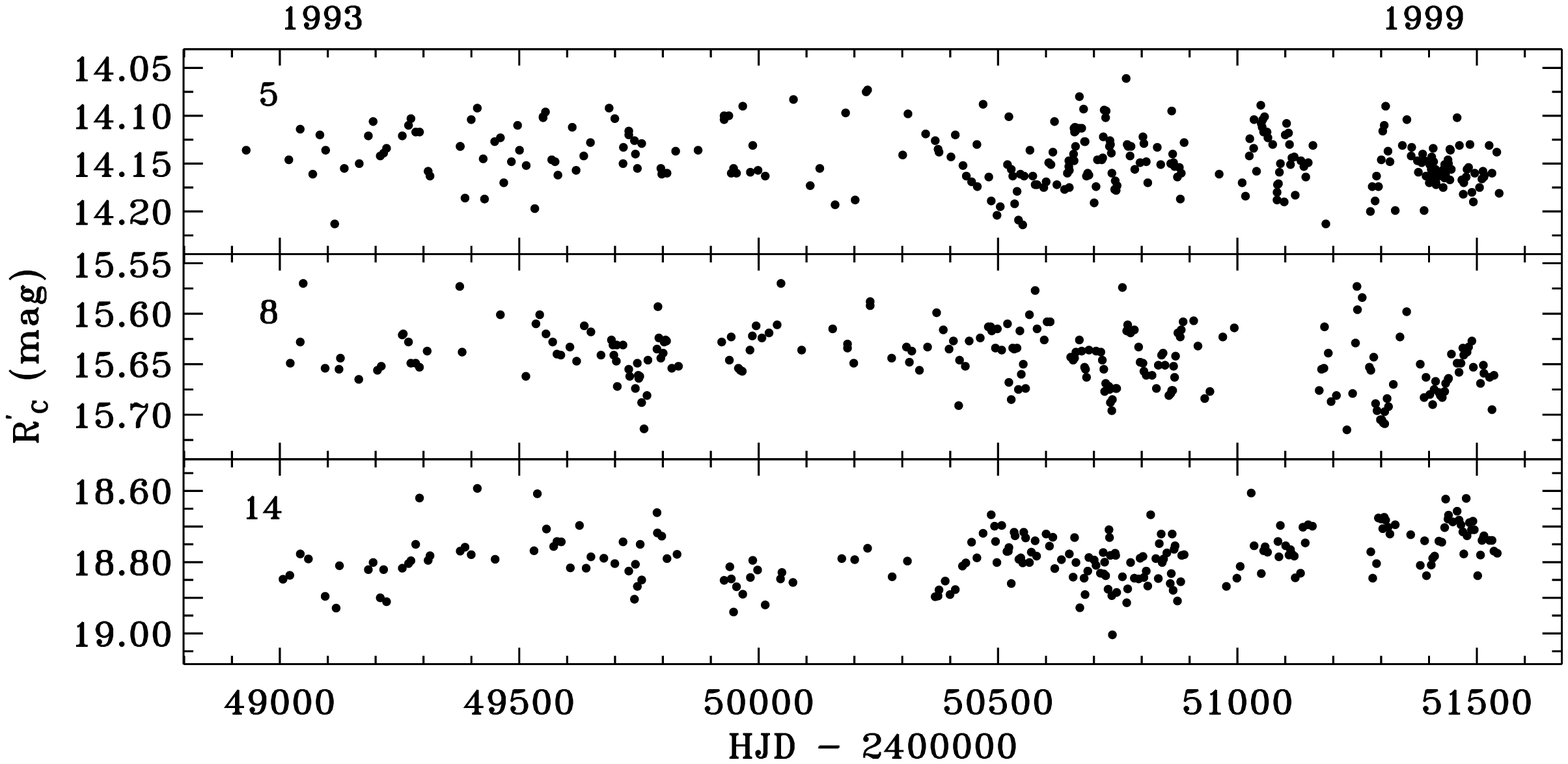}
\caption{The {\rm R}$^\prime_C$ light curves for the three objects with MACHO but not OGLE data.  The numbers in the upper left of each panel refer to the object identifications listed in Table~\ref{object_list}.  A zero-point correction was added to the instrumental MACHO red magnitudes to approximately bring them to the standard system.
\label{fig3}}
\epsscale{1.0}
\end{figure}

\begin{figure}\figurenum{4}\epsscale{0.85} 
\plotone{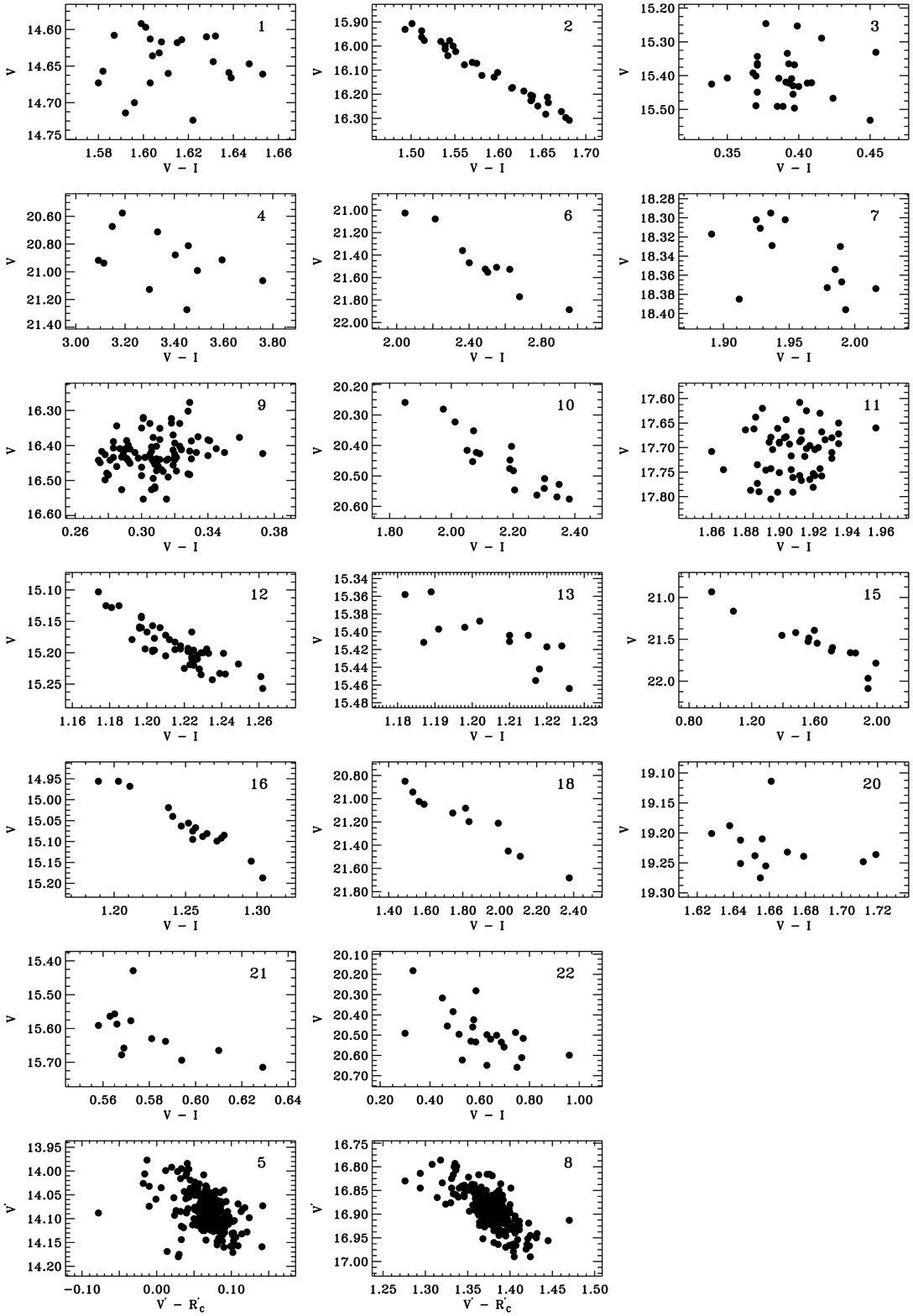}
\caption{The color curves for the target objects: ({\it V$-$I}) versus {\it V} for the OGLE data and 
({\rm V$^{\prime}$}$-${\rm R}$^\prime_C$) 
versus {\rm V$^{\prime}$} for the MACHO data.
\label{color}}
\epsscale{1.0}
\end{figure}

\begin{figure}\figurenum{5}\epsscale{0.90} 
\plotone{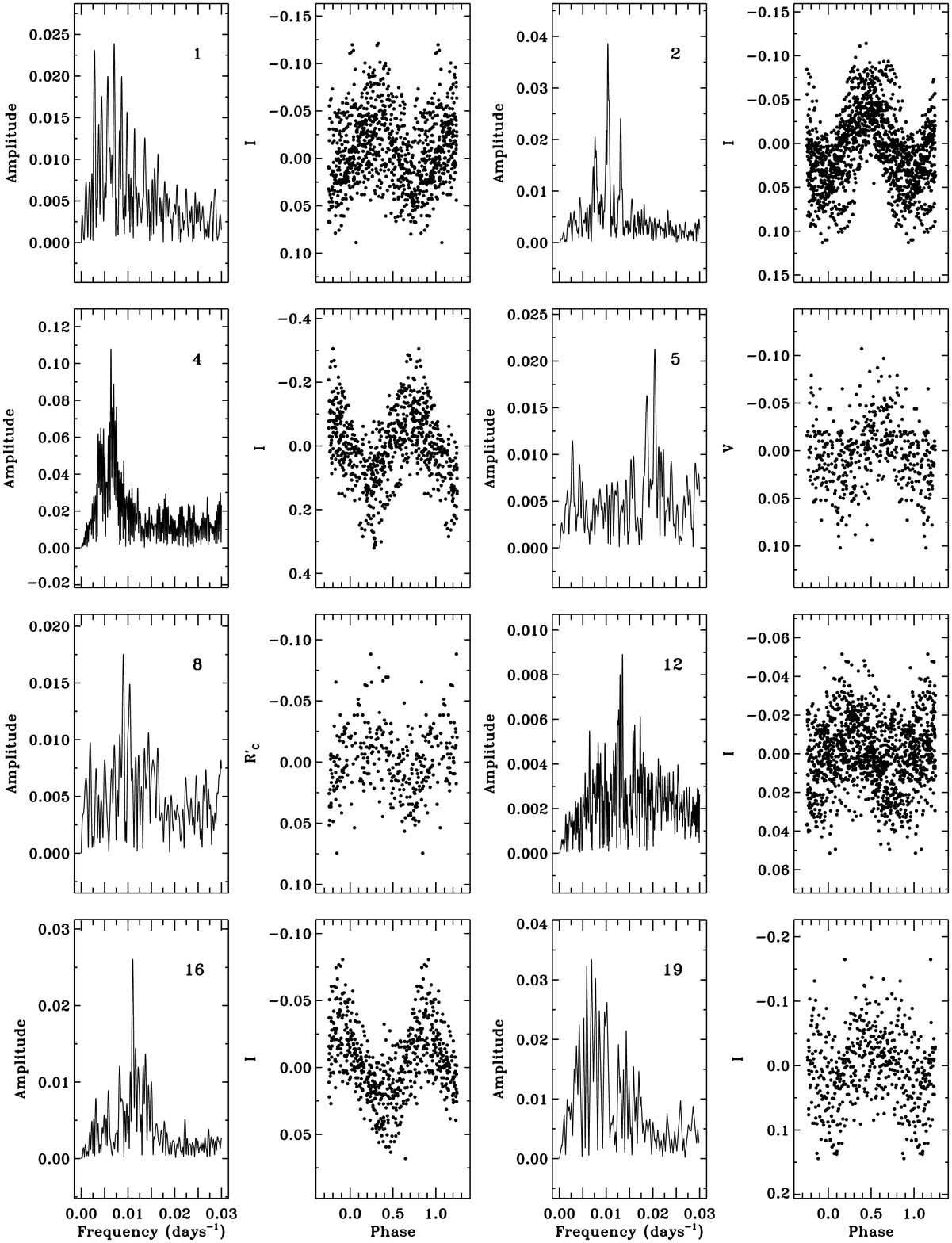}
\caption{ The frequency spectrum and the seasonally-normalized observations, phased with their dominant periods, for the eight periodic objects.  
\label{Pspec}}
\epsscale{1.0}
\end{figure}

\begin{figure}\figurenum{6}\epsscale{1.10} 
\plotone{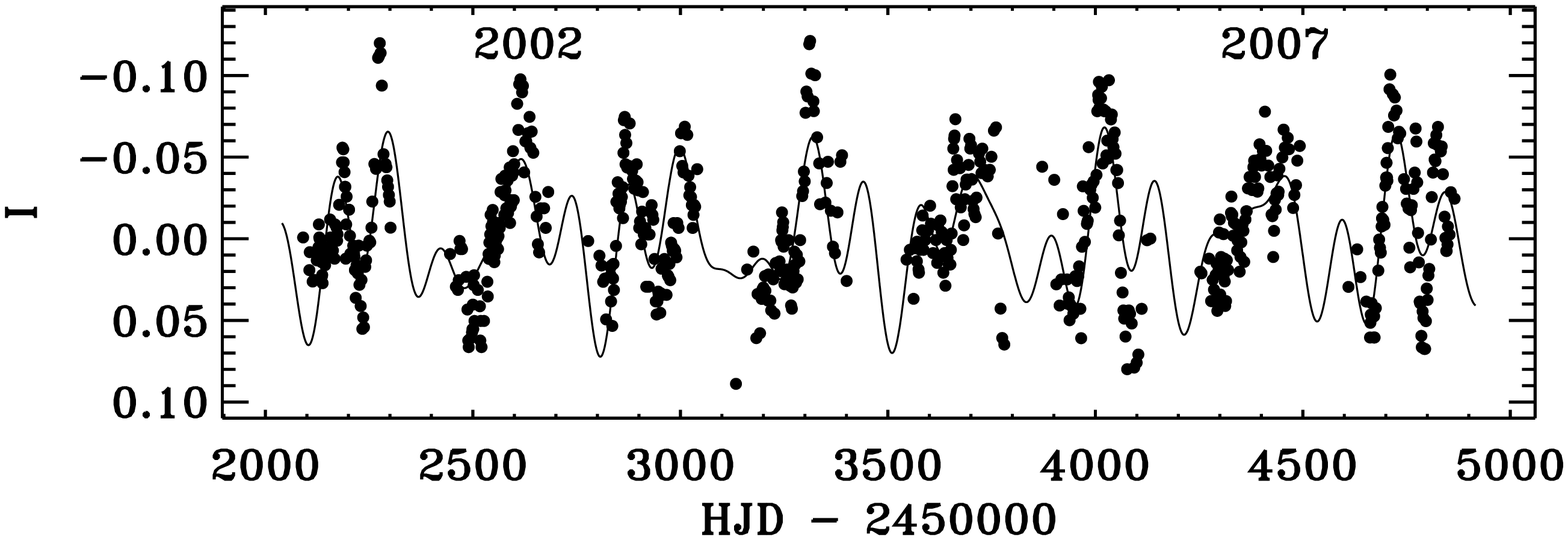}
\caption{The normalized {\it I} light curve of object 1 (J053250.69-713925.8), fitted by the three periods and amplitudes listed in  Table~\ref{periods}.
\label{obj20n_LC}}
\epsscale{1.0}
\end{figure}

\begin{figure}\figurenum{7}\epsscale{1.00} 
\plotone{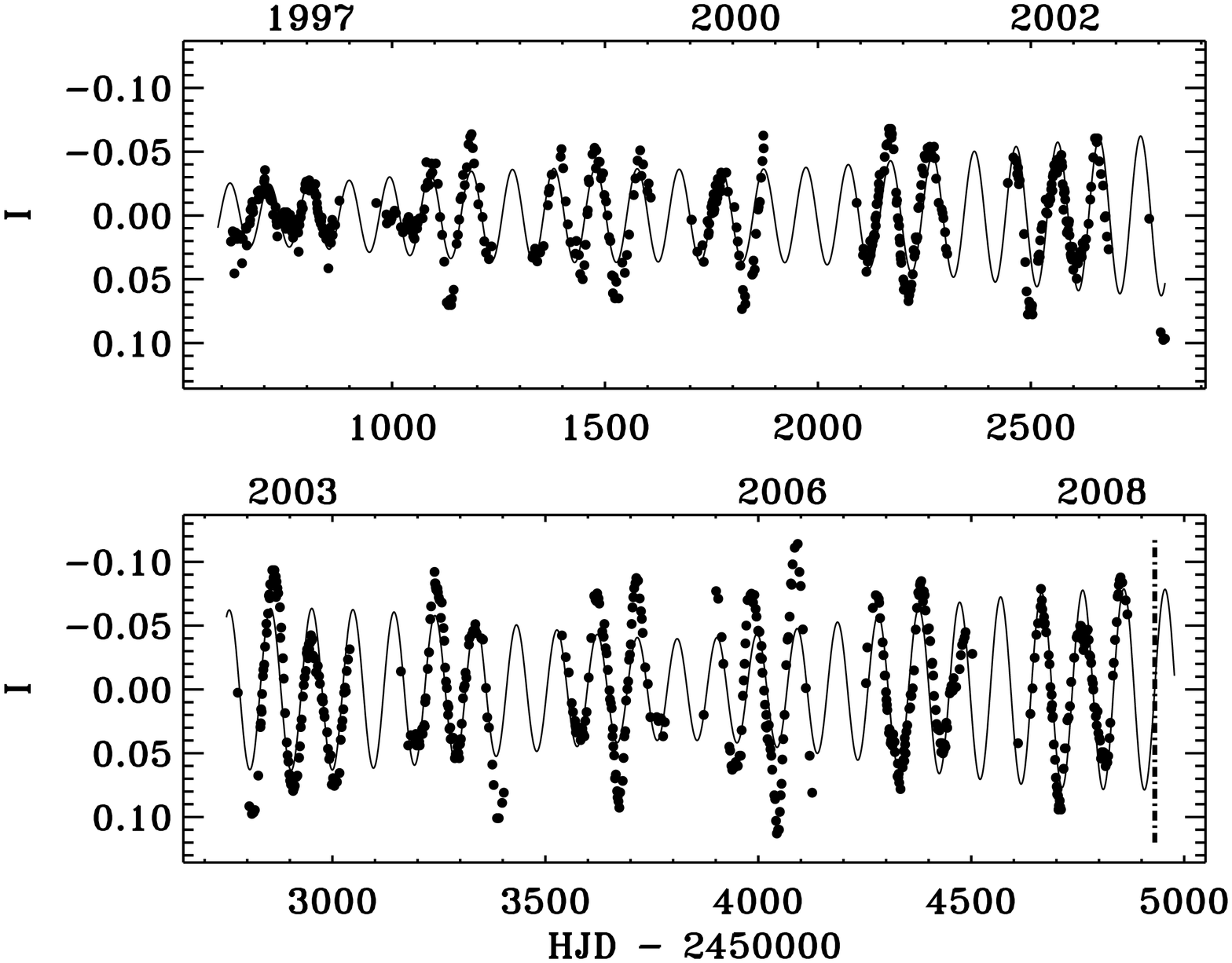}
\caption{The seasonally-normalized {\it I} light curve of object 2 (J004441.03-732136.44), fitted by the three periods and amplitudes listed in  Table~\ref{periods}.
The dot-dash vertical line indicates the time (and phase) of the published high-resolution spectrum.
\label{obj1_LC}}
\epsscale{1.0}
\end{figure}

\begin{figure}\figurenum{8}\epsscale{1.10} 
\plotone{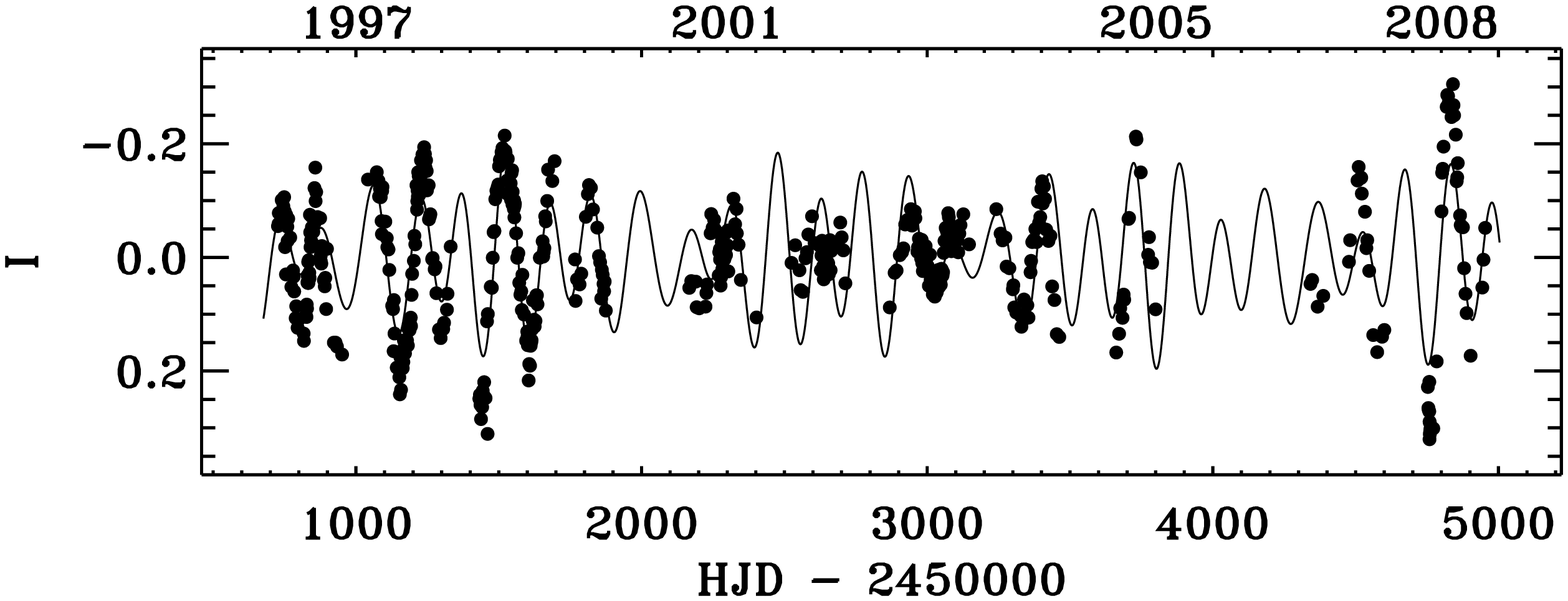}
\caption{The normalized {\it I} light curve of object 4 (J050603.66-690358.9), fitted by the three periods and amplitudes listed in  Table~\ref{periods}.
\label{obj3_LC}}
\epsscale{1.0}
\end{figure}

\begin{figure}\figurenum{9}\epsscale{1.10} 
\plotone{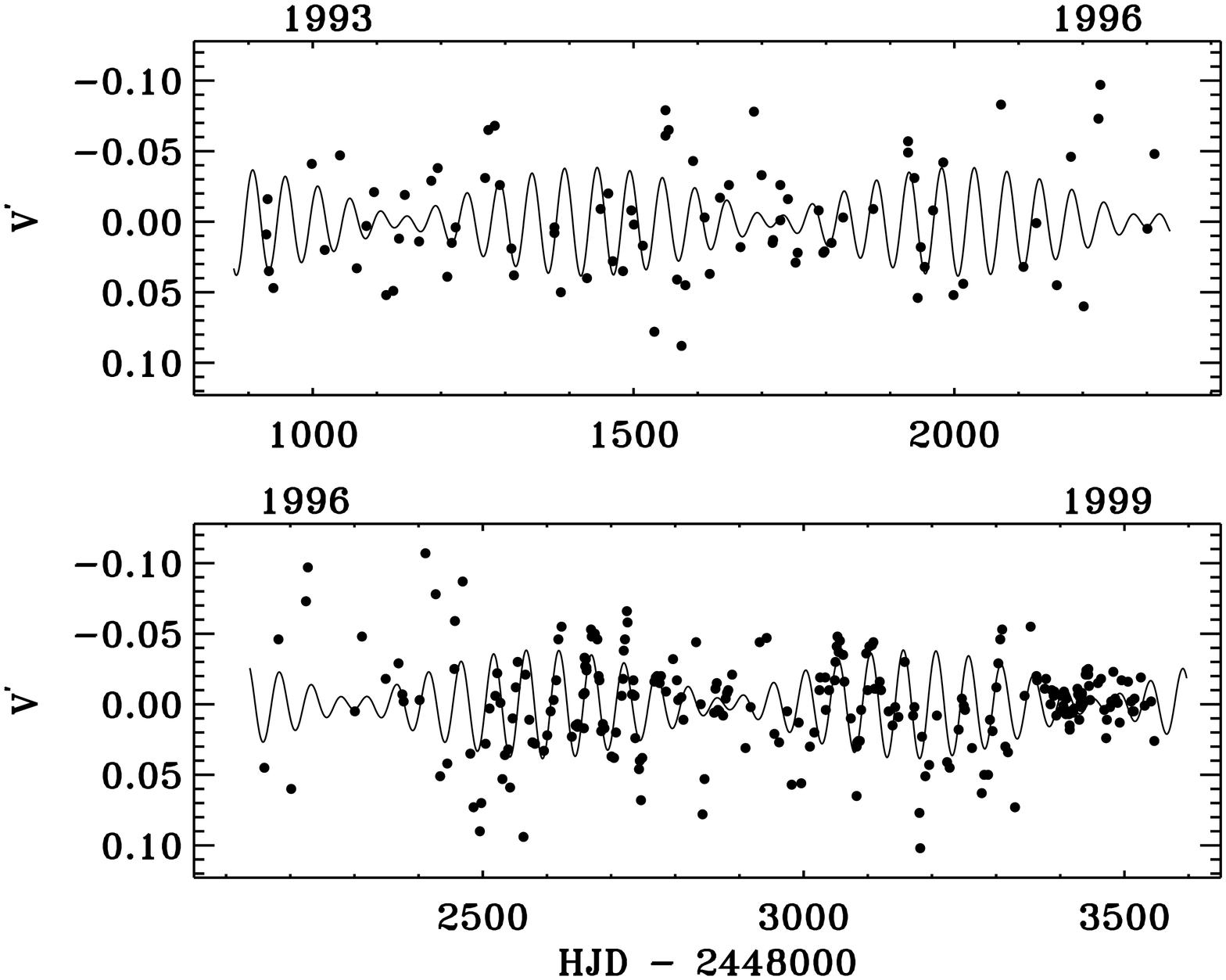}
\caption{The normalized {\rm V}$^{\prime}$ light curve of object 5 (J050632.10-714229.8) with the decreasing trend removed, fitted by the two periods and amplitudes listed in  Table~\ref{periods}.
\label{obj19_LC}}
\epsscale{1.0}
\end{figure}

\begin{figure}\figurenum{10}\epsscale{1.10} 
\plotone{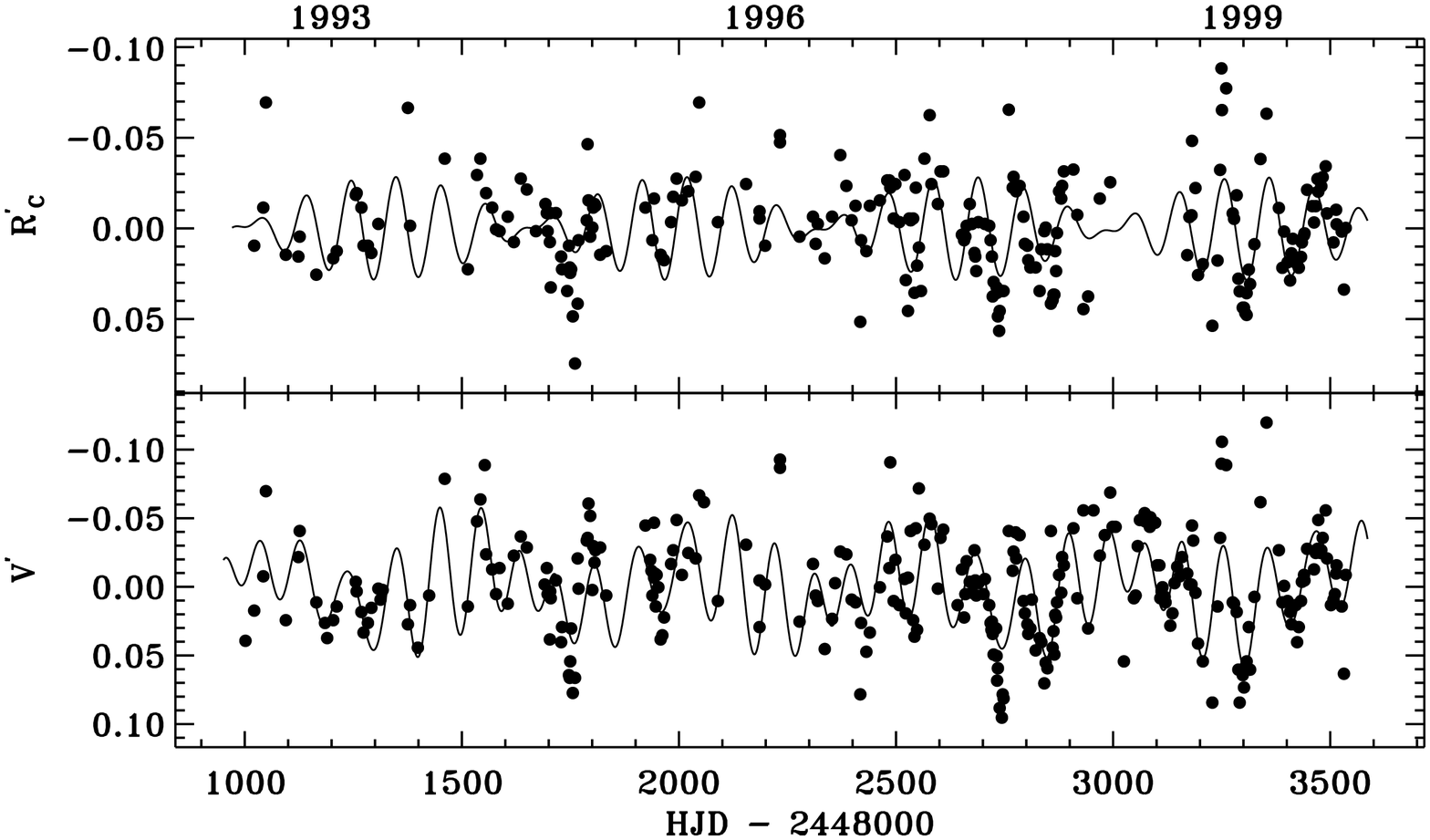}
\caption{The normalized {\rm R}$^{\prime}_C$ and {\rm V}$^{\prime}$ light curves of object 8 (J051110.64-661253.7), fitted by the periods and amplitudes listed in  Table~\ref{periods}.
\label{obj6_LC}}
\epsscale{1.0}
\end{figure}

\begin{figure}\figurenum{11}\epsscale{1.10} 
\plotone{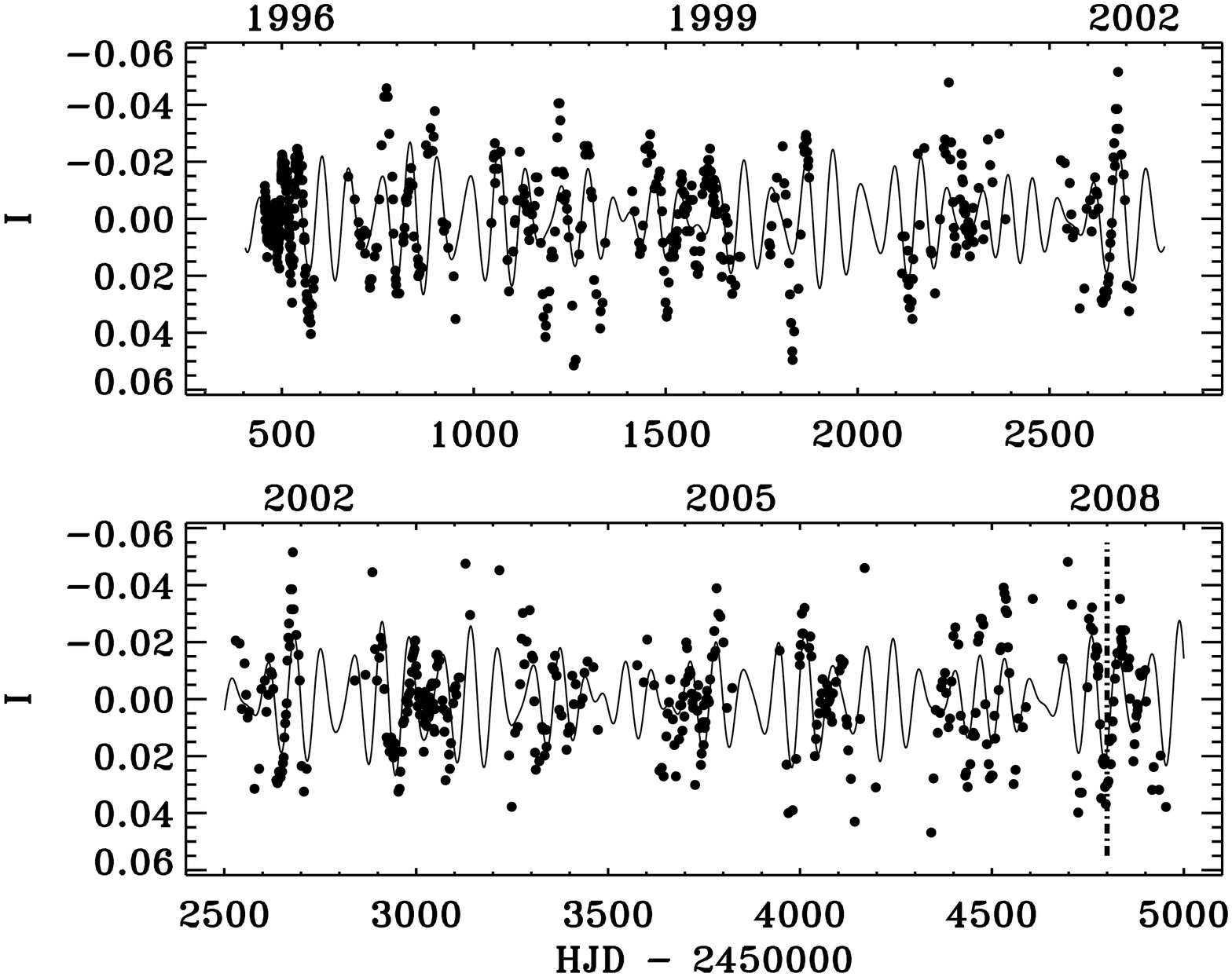}
\caption{The normalized {\it I} light curve of object 12 (J052043.86-692341.0), fitted by the four periods and amplitudes listed in  Table~\ref{periods}.
The dot-dash vertical line indicates the approximate time (and phase) of the published high-resolution spectra.
\label{obj10_LC}}
\epsscale{1.0}
\end{figure}

\begin{figure}\figurenum{12}\epsscale{1.10} 
\plotone{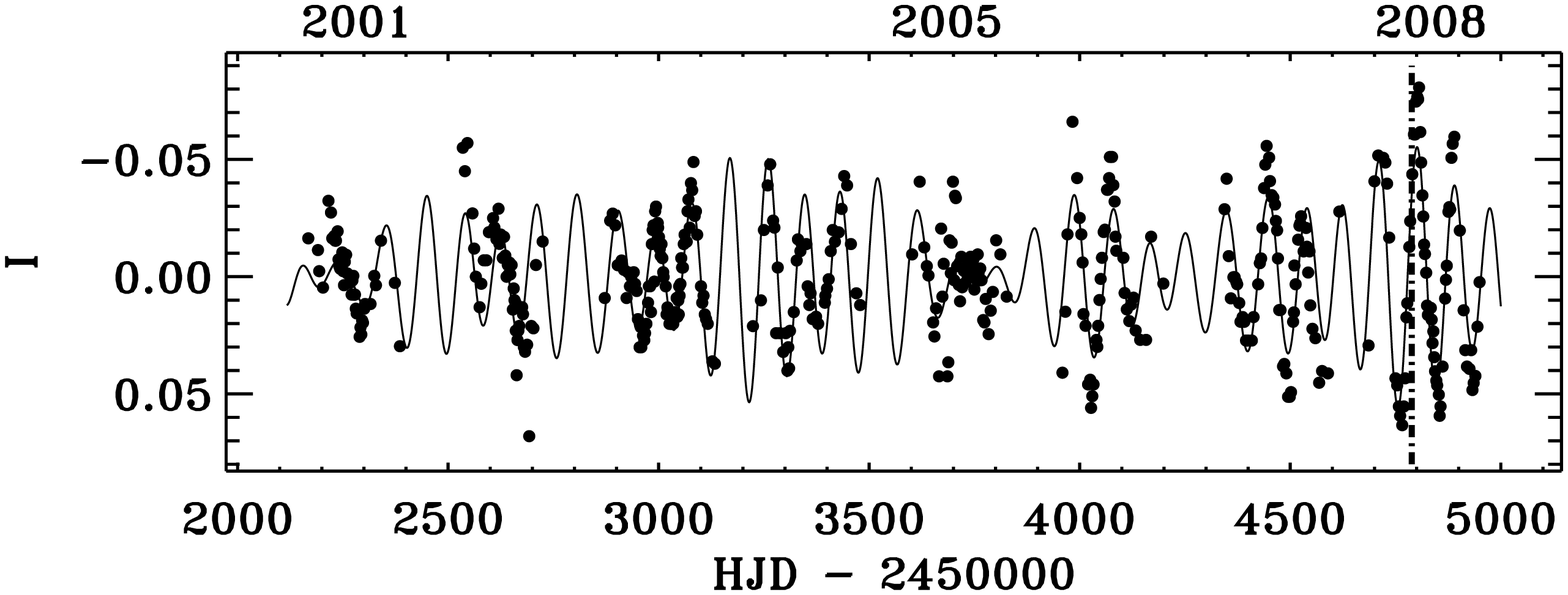}
\caption{The normalized {\it I} light curve of object 16 (J053250.69-713925.8), fitted by the four periods and amplitudes listed in  Table~\ref{periods}
The dot-dash vertical line indicates the time (and phase) of the published high-resolution spectrum.
\label{obj22_LC}}
\epsscale{1.0}
\end{figure}

\begin{figure}\figurenum{13}\epsscale{1.10} 
\plotone{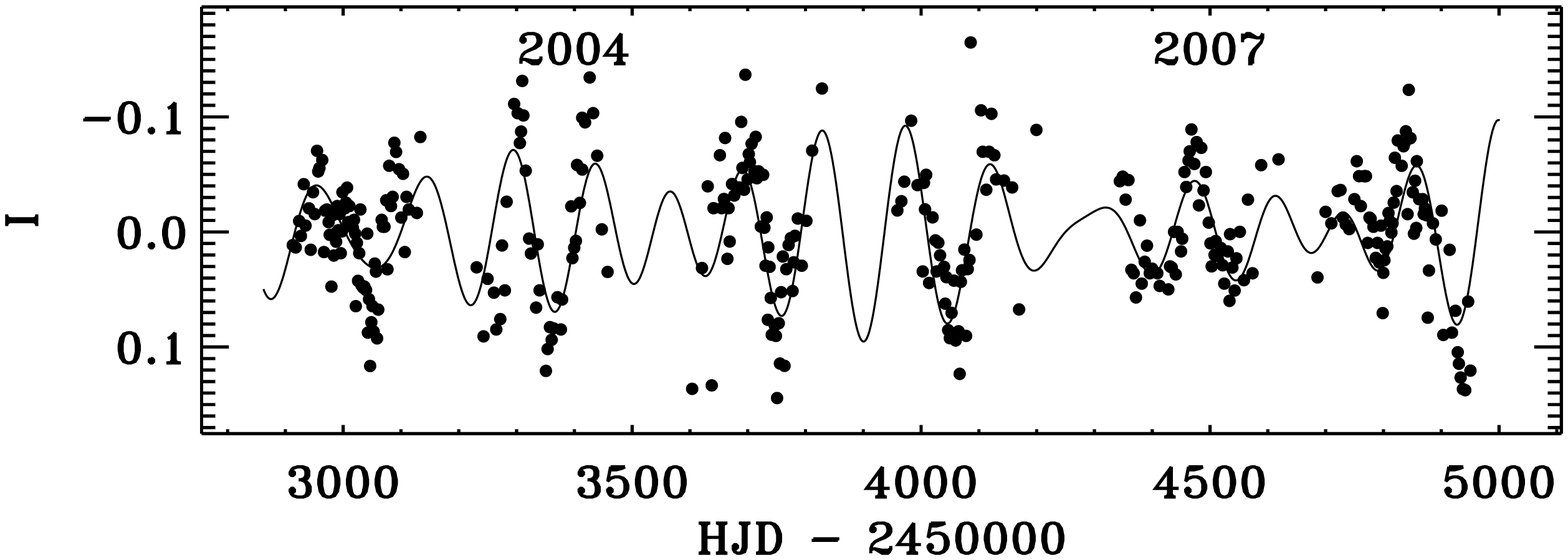}
\caption{The normalized {\it I} light curve of object 19 (J054054.31-693318.5), fitted by the three periods and amplitudes listed in  Table~\ref{periods}.
\label{obj22n_LC}}
\epsscale{1.0}
\end{figure}

(15) {\it J053044.10-714300.6} (Figure~\ref{fig2b}):
This object is faint and red, with $<${\it I}$>$ = 19.92, $<$$\sigma$$_I$$>$ = 0.13, and $<$({\it V$-$I})$>$ = 1.62 mag.  
The light curve shows a relatively constant seasonal mean brightness over 8 years, with the first year a bit fainter.
There is a relatively large range in brightness within a season, with an average variation of 0.71 mag ({\it I}) and a range in seasonal variations of 0.57 to 0.92 mag.
The variations are short-term, of a few days or less, with no apparent cyclical behavior.  No periodicity is found in the data.   
It shows a very large variation in color, $\Delta$({\it V$-$I}) $\approx$ 1.0, although most of the data points (13/15) are within a range of 0.6 mag, and a clear trend of being redder when fainter (Figure~\ref{color}).
	
(17) {\it J053525.86-711956.6} (Figure~\ref{fig2b}): 
The average value of {\it I} is faint at $<${\it I}$>$ = 18.96 and the uncertainties in the data are large, with $<$$\sigma$$_I$$>$ = 0.05.  The light curve shows a monotonic increase in brightness over the first six seasons of 0.19 mag and then a decline of 0.09 mag, which could be fit with a long-term ($\sim$10 year) cyclical variation.  The variations within a season range from 0.26 to 0.38 mag and are very short term, a few days or less, with no apparent cyclical behavior; no periodicity was found.  There are no OGLE {\it V} observations of this object.
	
(18) {\it J053632.56-701738.4} (Figure~\ref{fig1}):
This object has both OGLE-II and III data.  It is faint and we removed some data with large uncertainties 
($\sigma$$_I$$>$0.125).
The average value of {\it I} is 19.39 mag and the uncertainties in the data are large, with $<$$\sigma$$_I$$>$ = 0.08 mag. 
The seasonal mean brightness do not vary much from 1999 to 2008, but they are fainter by 0.11 mag (I) for the first two seasons, and within a season the light varies by $\sim$0.5 mag ({\it I}). 
The variations are very short term, a few days or less, with no apparent cyclical behavior, and no periodicity was found.  The object is faint in {\it V}, with average values of $<${\it V}$>$ = 21.2 mag and red, with $<${\it V$-$I}$>$ = 1.8 mag. It varies a lot in color, 
$<$$\Delta${\it (V$-$I)}$>$ = 0.9 mag, and is redder when fainter (Figure~\ref{color}).
	
(20) {\it J055311.98-701522.7} (Figure~\ref{fig2b}):
The light curve shows a gradual increase in seasonal mean brightness of 0.08 mag ({\it I}) over 8 years.  The average brightness variation is $\sim$0.17 mag in a season; the variations are short-term of a few days or less, with no apparent cyclical behavior.  No periodicity is found in the data.  The average brightness is $<${\it I}$>$ = 17.60
and it is red, with $<${\it V$-$I}$>$ = 1.66 mag. No correlation is found between brightness and color (Figure~\ref{color}). 
	
(21) {\it J055825.96-694425.8} (Figure~\ref{fig2b}):
The object is very bright, with $<${\it I}$>$ = 15.04, 
and is relatively blue, with $<$({\it V$-$I})$>$ = 0.58.  The light curve shows a decrease in mean brightness from the first season to the second by $\sim$0.06 mag and from the second to the third by $\sim$0.04 mag, but then is relatively constant in mean brightness, with the next to last year (2007) increasing slightly from this mean and the last year (2008) decreasing slightly.  The range in a season is $\sim$0.25 mag.  There are suggestions of cyclical variations on the order of 10 days in the data, which may be due partly to the sampling frequency, but no periodicity is found.  There appears to be a general correlation between faintness and redness in the color (Figure~\ref{color}).
	
(22) {\it F06111-7023} (Figure~\ref{fig2b}):	
This object is faint and the observations possesses large uncertainties, with values of $<${\it I}$>$ = 19.89
and with $<$$\sigma$$_I$$>$ = 0.13.  
It was observed in two OGLE-III fields and thus has two OGLE-III identifications.  Taken individually, LMC207.3\_5393 shows a slight increase in brightness of $\sim$0.07 mag ({\it I}) in the average seasonal light levels over the eight years of observations, while within a season most of the data vary within a range of 0.35 mag ({\it I}).  This is the light curve that we have plotted.
In the second field, where it is identified as LMC214.6\_915, there is an apparent increase in brightness by $\sim$0.20 mag ({\it I}) over the eight years, while within a season, most of the data vary within a range of 0.5 mag ({\it I}).  In both fields, the object appears to show real variations of a short timescale of $\sim$8 days or less, with the upper limit set partly by the sampling frequency, with no apparent cyclical variation.  No periodicity is found in the data.  When we combine the two data sets, they show an average increase in brightness of 0.13 mag ({\it I}) over the eight years of observations.  We normalized them separately and then combined them for the period search.  The data set then has many pairs of observations made on the same day that often vary by several tenths of a magnitude, suggesting a very short timescale of variation.  
We again find no periodicity, but suggestions of a timescale for the variability of a few days. 
The comparison of the data for this star from the two separate fields shows that one must regard with caution the trends in the data for such a faint object with large uncertainties in the data.  
The object is relatively blue, with with $<${\it V$-$I}$>$ = 0.60, but varies over a large range in color,
$\Delta$({\it V$-$I}) = 0.7, and appears in general to be redder when fainter (Figure~\ref{color}).

\section{APPENDIX II: DETAILS OF THE NEW SPECTRA}
\label{Spec-details}

\subsection{Observations and Reductions}

Spectra of the following five objects were obtained at Gemini South using GMOS with the R400 grating, central wavelengths of 6000 and 6050 \AA, and a slit width of 0.75\arcsec.
 Either 4 or 8 individual exposures of 450 seconds each were taken for the different objects, depending on their optical brightnesses, and these spectra were combined to make the final spectrum for each object. 
To allow relative flux calibration of the spectra, the spectrophotometric standard Hiltner 600 was observed in the same configuration, although on a different night. 
The standard data reduction was done using the Gemini IRAF package. 
Note that the Gemini baseline calibrations allow relative spectral calibration but do not allow an accurate absolute flux calibration of the spectrum.  Comparison of the 
nominal flux levels of the target spectra to the optical photometry of the objects shows that the spectra are generally 
a factor of 2 to 3 times brighter than expected, which probably is due to significant slit losses in the 
spectrophotometric standard spectra, which were taken during clear conditions but with poor seeing.

There is significant fringing in the long wavelength end of the spectra. This was removed by use of the standard star spectrum. We first fit and removed the stellar spectral lines from the standard star spectrum, and then carried out continuum fitting and normalization of the resulting $\arcsec$featureless$\arcsec$ spectrum. This normalized spectrum was rebined to match the target spectrum wavelength sampling to correct for the small zero-point shift between the two spectra. 
Finally we divided the extracted target spectrum by the rectified spectrum. This process was found to remove much of the fringing that is seen in the spectra at $\lambda$ $>$ 7000 \AA, and it also removed some fraction of the atmospheric bands at 6860 \AA~ and 7600 \AA. Since the standard star spectrum was taken at a much lower airmass than the science observations and 1 to 3 months earlier, no atmospheric extinction correction was attempted. In any case the main concern was correcting the spectral fringing rather than removing the telluric bands.

\subsection{SAGE J051110.64$-$661253.7 (no. 8) and J052043.86$-$692341.0 (no. 2)}

Based upon the G-band/H$\gamma$ ratio and the H$\alpha$/H$\beta$ line strengths, object 12 is classified as type F8 I(e). The luminosity class is uncertain as the standard luminosity criteria in the blue part of the optical spectrum could not be applied due to the large reddening of the spectrum.  
The luminosity is class I or II based upon the general line widths in the spectrum as compared with the spectra from \citet{jac84}, which have very similar resolution. The spectral type also has an uncertainty because the Balmer lines are weak and would suggest a later spectral type of about G2, 
whereas for that type of spectrum the G-band would be expected to be much stronger than is observed. This probably means that the Balmer lines are partially filled in by emission, although there is no sign of core reversal in the line profiles; hence we have added the (e) qualifier to the spectral 
classification. The spectral type could be a little earlier than F8 if the photosphere is carbon-rich (as implied by the mid-infrared spectrum) and the G-band is thereby enhanced.

For object 8, the spectrum declines even more sharply in the blue part of the optical than is the case for object 12, and this makes classification even more difficult. At the same time, the spectrum does not extend far enough to long wavelengths to use the infrared Ca triplet and the Paschen 
series lines for classification. As a result we have been unable to assign a definite spectral type to the star using the Gemini observations. The star is classified as type F3 II(e) by \citet{aarle11}. From our spectrum we find that the 
Balmer lines are definitely weaker than would be expected for an early F-type star, being more like the strengths expected for an early G-type star.  However, the G-band is not detected so the star cannot be as late as the Balmer line strengths would suggest. This is all consistent with the 
F3 II(e) classification. 

\subsection{SAGE J051228.18$-$690755.8 (no. 9)}

Object 9 shows strong H$\alpha$ emission along with the [N II] $\lambda\lambda$6548/6584 \AA~ nebular lines on a relatively blue continuum. The H$\beta$ line is also seen in emission. The
H$\alpha$/H$\beta$ line flux ratio is 4.3, which is much higher than the normal value of 2.85, presumably due to the extinction from the dust shell. The c(H$\beta$) extinction value is estimated to be 
0.56 on this basis. The line fluxes may be affected by an underlying absorption component, but there is no sign of this in the spectrum. There seems to be a broad, low-level emission component extending 
from 6505 \AA~ to 6608 \AA~ under these emission lines. If this is due to H$\alpha$ emission wings from a stellar wind, it implies a wind speed of almost 3000 km~sec$^{-1}$.

All the absorption lines in the spectrum are weak compared to the standard star spectra. The H$\gamma$ line is probably present in weak emission, while the H$\delta$ line is detected weakly in absorption. 
A few He I lines are present, and the He II lines such as $\lambda\lambda$4686 \AA~ and $\lambda\lambda$4541 \AA~ are absent. There are also absorption lines of O II, N III and C III visible in the spectrum. 
Based upon comparison of these few lines with spectral standards it appears that the spectral type is B0.5 I[e], with the luminosity class being assigned on the basis of the narrowness of the absorption lines 
of He, C, N, and O that can be identified, since the normal luminosity indices involve lines that are are not detected in the spectrum.

It is possible that the spectral type is assigned too early a value if the stellar C abundance is significantly enhanced, since the strength of the C~III line at $\lambda\lambda$4650 \AA~ compared to other nearby lines is significant in assigning the B0.5 spectral type. However it seems unlikely to be later than about B2 given the lines detected in the spectrum.
The LSR radial velocity is found to be 175 km~s$^{-1}$ from the emission lines (H$\alpha$, H$\beta$,
[N~II] $\lambda\lambda$6584 \AA), which is lower than for the two field stars that also were observed on the slit (see next section). This indicates that even though the object appears to be in the LMC bar, it is rather an LMC halo star in the foreground or background.\\

\subsection{FIELD STARS}

\begin{figure}\figurenum{14}\epsscale{1.0} 
\rotatebox{0}{\plotone{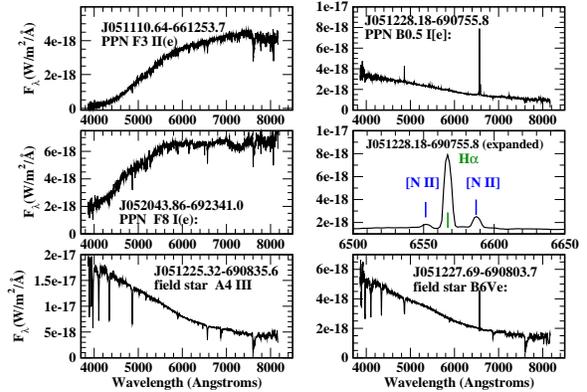}}
\caption{Optical spectra obtained for three of the targets along with two field stars. 
Objects 8 (J051110.64$-$661253.7) and 12 (J052043.86$-$692341.0)
have heavily reddened spectra. The spectral type for the first of these objects cannot be obtained 
from the spectrum, and the F3 II(e) classification is from \citet{aarle11}.
We have classified the other two, along with two field stars; 
these later show the quality of the spectroscopy for normal stars.
For object 9 (J051228.18$-$690755.7),  the region around H$\alpha$ is plotted in 
expanded scale to show the [N II] nebular forbidden line emission. 
\label{spectra}}
\epsscale{1.0}
\end{figure}

\begin{figure}\figurenum{15}\epsscale{0.80} 
\rotatebox{90}{\plotone{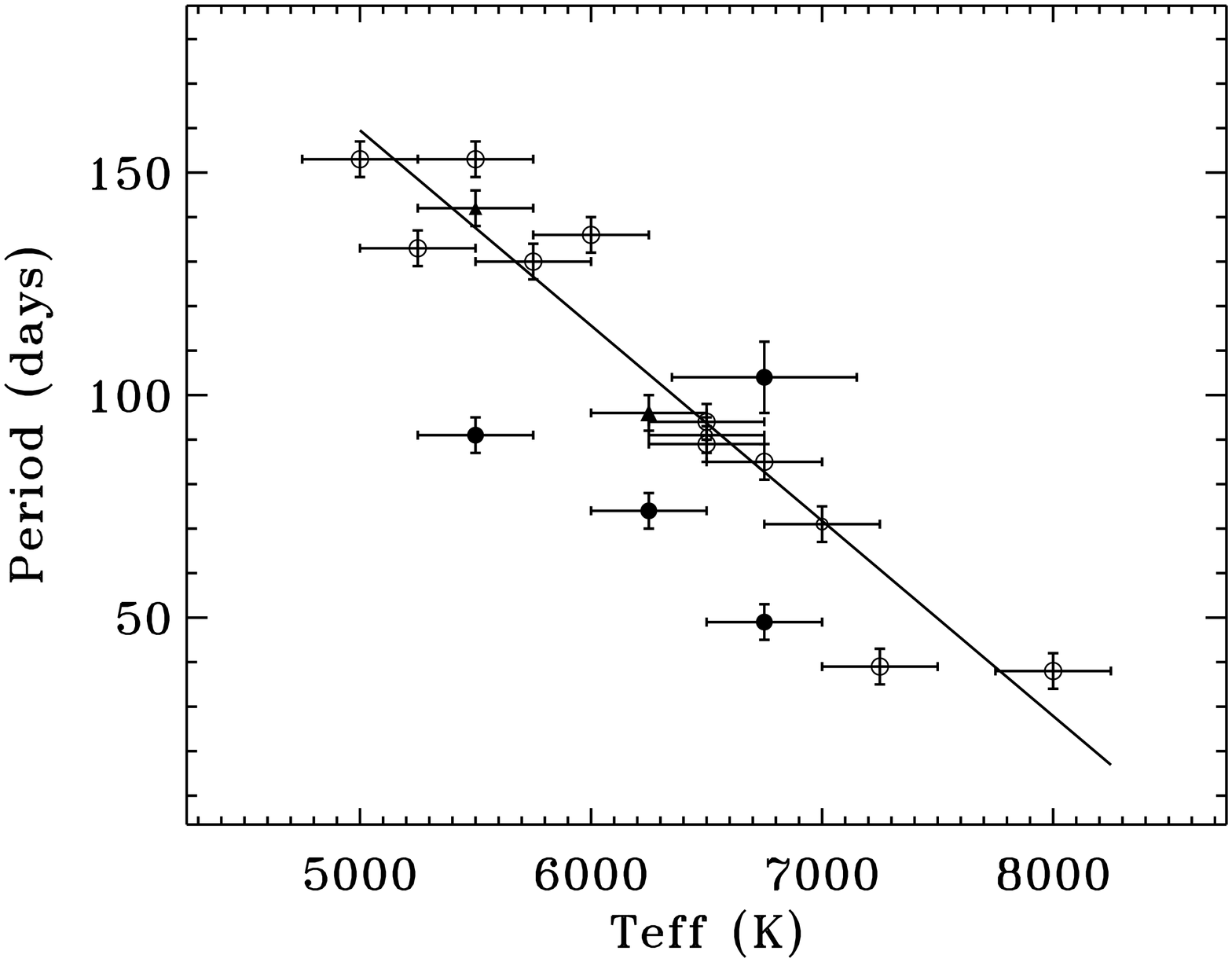}}
\caption{A plot of pulsation period versus T$_{\rm eff}$.  The 12 Milky Way objects are shown with open circles \citep{hri10} and the five new Megallanic Cloud objects are shown with filled circles (4, LMC) or a filled triangle (1, SMC). 
Those with less certain values are shown with smaller symbols. 
For object 8, we have used the average period.
The solid line is a linear fit to the distribution of values for the Milky Way objects.
\label{P-T}}
\epsscale{1.0}
\end{figure}

\begin{figure}\figurenum{16}\epsscale{0.80} 
\rotatebox{90}{\plotone{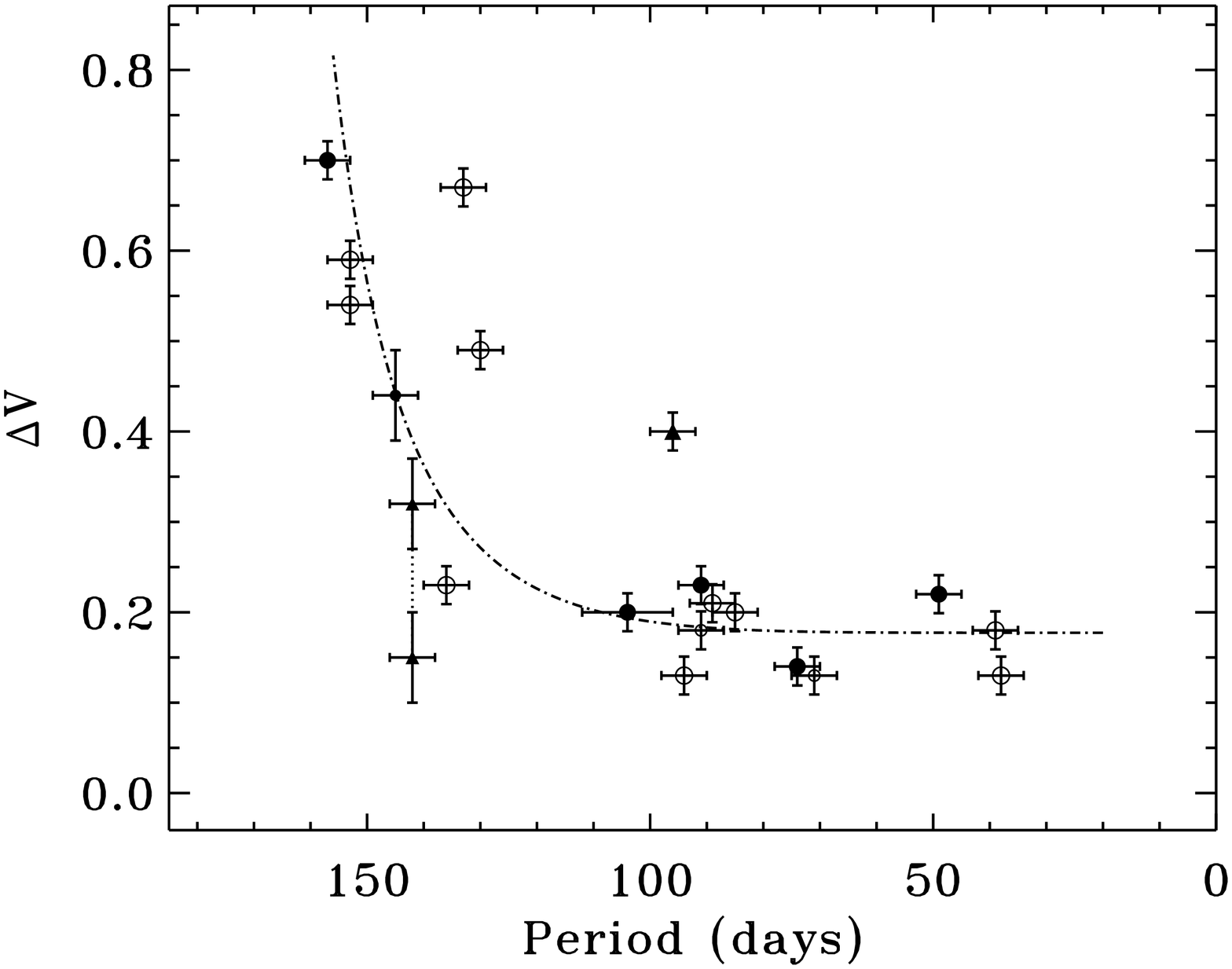}}
\caption{A plot of pulsation period versus maximum {\it V} amplitude in a season.  The 12 Milky Way objects are shown with open circles \citep{hri10} and the new Megallanic Cloud objects are shown with filled circles (LMC) or filled triangles (SMC). 
Those with less certain values are shown with smaller symbols. 
For object 8, we have used the average period.
The vertical dotted line connects the observed and scaled maximum {\it V} amplitude for object 1; for object 19, which does not have {\it V} observations, we plotted only the scaled maximum {\it V} amplitude.  
Since the {\it V} amplitudes for objects 1 and 19 are less certain, they are assigned larger error bars.
The dot-dash line is an approximate fit to the distribution of values for the Milky Way objects.
\label{P-Amp}}
\epsscale{1.0}
\end{figure}

In the observation of Object 9, the slit was oriented 21$\arcdeg$ E of N to also allow spectra to be taken of two other objects in the field. These two additional objects are $\sim$15$\arcsec$ and $\sim$49$\arcsec$ away from the PPN candidate to the south and west. The spectra of these two objects were reduced and are also presented in Figure~\ref{spectra}. 
For these two objects, being located at much different positions along the slit than the main targets, the correction for the fringing using the standard star spectrum tended to increase the residual fringing noise at long
wavelengths rather than suppress it, and so the presented spectra are not corrected in this
manner. Otherwise the reduction was the same for these spectra as for the target spectra.

The brighter of these two additional objects is USNO-B 0208-0081531 = 2MASS
J05122533$-$6908355 = SAGE J051225.29$-$690835.6.
Based on the 2MASS and SAGE
photometry for this object it was expected to be a A-type giant or supergiant in the LMC. 
The second object is 2MASS J0512768$-$6908039 = SAGE J051227.69$-$690803.7. 
The main 2MASS catalogue has poor photometry for this second object compared to the 2MASS
6X 
catalogue, where the object position is given as R.A. = 05$^h$12$^m$27.68$^s$ and Decl. =
$-$69$\arcdeg$08$\arcmin$03.68$\arcsec$ (J2000). 
The combined 2MASS/SAGE photometry for this object suggests that it is a B-type star in the LMC.

Object SAGE J051225.29$-$690835.6 is classified spectroscopically as an A-type giant star. A spectral
type of A4 III is assigned based upon the Balmer series line profiles, the lack of He lines,
and the Ca II K to H$\epsilon$ line ratio. From the Balmer lines, the radial velocity V$_{LSR}$ is about
215 km~s$^{-1}$, somewhat lower than the nominal value of V$_{LSR}$ = 250$\pm$5 km km~s$^{-1}$ for the LMC, based on  radial velocities of PNe \citep{meath88}.  

Object SAGE J051227.69$-$690803.68 is classified as a main sequence B-type star with
H$\alpha$ emission. A spectral type of B6~Ve is assigned based on the Balmer lines, the Ca~II K 
$\lambda\lambda$3933 \AA~ line, and the Si~II $\lambda\lambda$4128-30/He I $\lambda\lambda$4144 
line ratio.
The radial velocity measured from the H$\beta$ and H$\gamma$ lines is V$_{LSR}$ of 220 km~s$^{-1}$. 
In both cases the spectroscopy verifies the expected spectral types based upon the 1 to 5 $\mu$m photometry.

\clearpage

\begin{deluxetable}{rrrrrrrr}
\rotate \tablenum{1}
\tablecolumns{8} \tabletypesize{\scriptsize}
\tablecaption{Target ID\tablenotemark{a} 
\label{object_list}} 
\tablehead{
 \colhead{No.} & \colhead{SAGE ID} & \colhead{2MASS ID} & \colhead{IRAS ID} & \colhead{OGLE-III ID} & \colhead{OGLE-II ID} & \colhead{OGLE VAR ID} & \colhead{MACHO ID\tablenotemark{b}} 
}\startdata
1 & J003659.53-741950.4 & J00365957-7419503 & 00350-7436 & SMC134.4.3525 & \nodata & SMC$-$LPV$-$02074 & \nodata \\
2 & J004441.03-732136.44 & J00444111-7321361 & \nodata & SMC125.2.42447 & SMC\_SC3\_130213 & SMC$-$LPV$-$04910 & \nodata \\
3 & J010546.40-714705.27 & J01054645-7147053 & \nodata  & SMC114.1.65  & \nodata & \nodata & \nodata\\
4 & J050603.66-690358.9 & J05060367-6903587 & 05063-6908  & LMC119.4.231 & LMC\_SC13\_79398 & \nodata & \nodata \\
5 & J050632.10-714229.8\tablenotemark{d} & J05063210-7142298 & \nodata & \nodata & \nodata & \nodata & 38.4253.12 \\
6 & J050713.90-674846.7 & J05071393-6748467 & 05073-6752 & LMC117.4.30028 & \nodata & \nodata & \nodata\\
7 & J050835.91-711730.6 & J05083592-7117306 & 05092-7121 & LMC114.8.182 & \nodata & \nodata & \nodata\\
8 & J051110.64-661253.7 & J05111065-6612537 & F05110-6616  & \nodata & \nodata & \nodata & 56.5062.17 \\
9 & J051228.18-690755.8 & J05122821-6907556 & 05127-6911 & LMC111.6.76310 & LMC\_SC10\_255396 & \nodata & \nodata \\
10 & J051828.18-680404.0 & J05182817-6804040 & F05186-6807  & LMC102.7.76310 & \nodata & \nodata & \nodata\\
11 & J051845.23-700534.5 & J05184525-7005344 & F05192-7008 & LMC103.8.24255 & \nodata & LMC$-$LPV$-$39671 & \nodata \\
12 & J052043.86-692341.0 & J05204385-6923403 & \nodata       & LMC100.2.18930 & LMC\_SC6\_205804 & LMC$-$LPV$-$43685 & \nodata \\
13 & J052520.76-705007.5 & J05252077-7050075 & Z05259-7052 & LMC164.5.20296 & \nodata & LMC$-$LPV$-$53645 & \nodata \\
14 & J052902.39-661527.7 & J05290241-6615277 & F05289-6617 & \nodata & \nodata & \nodata & 64.7965.525 \\
15 & J053044.10-714300.6 & \nodata & 05315-7145 & LMC172.7.7707 & \nodata & \nodata & \nodata \\
16 & J053250.69-713925.8\tablenotemark{d} & J05325071-7139257 & \nodata & LMC172.7.47076 & \nodata  & LMC$-$LPV$-$68005 & \nodata \\
17 & J053525.86-711956.6 & J05352583-7119566 & F05361-7121 & LMC172.4.47181 & \nodata & \nodata & \nodata\\
18 & J053632.56-701738.4 & \nodata & 05370-7019   & LMC177.6.60500 & LMC\_SC16\_166486 & \nodata & \nodata\\
19 & J054054.31-693318.5  & J05405428-6933185 & 05413-6934 & LMC176.4.40319 & \nodata & \nodata & \nodata \\
20 & J055311.98-701522.7 & J05531196-7015226 & 05537-7015 & LMC192.5.21143 & \nodata & \nodata & \nodata\\
21 & J055825.96-694425.8  & J05582596-6944257 & F05588-6944 & LMC198.6.5444 & \nodata & \nodata & \nodata\\
22 & \nodata  & J06103202-7024408 & F06111-7023 & LMC.207.3\_5393\tablenotemark{c} &\nodata & \nodata & \nodata\\
\enddata
\tablenotetext{a}{Additional catalog identifications for most are given by \citet[Table 2]{volk11}}
\tablenotetext{b}{MACHO ID listed for those not in the OGLE catalogs.}
\tablenotetext{c}{Also observed in a second OGLE-III field as LMC\_214.6\_915.  The two data sets are combined for use in this study.}
\tablenotetext{d}{From \citet{aarle11,aarle13}.}
\end{deluxetable}

\clearpage

\begin{deluxetable}{rlrrrrrrrrrr}
\rotate \tablenum{2}
\tablecolumns{12} \tabletypesize{\scriptsize}
\tablecaption{Observational Statistics\tablenotemark{a}
\label{statistics}} 
\tablehead{
\colhead{No.} & \colhead{SAGE ID} & \colhead{Years} 
& \colhead{No. ({\it I})} & \colhead{$<${\it I}$>$\tablenotemark{b}} & \colhead{$<$$\sigma$$_I$$>$\tablenotemark{c}} 
& \colhead{No. ({\it V})} & \colhead{$<${\it V}$>$\tablenotemark{b}} & \colhead{$<$$\sigma$$_V$$>$\tablenotemark{c}} 
& \colhead{No. ({\it V$-$I})} & \colhead{$<${\it V$-$I}$>$\tablenotemark{b}} 
& \colhead{$<$$\sigma$$_{V-I}$$>$\tablenotemark{c}}\\
\colhead{} & \colhead{} & \colhead{} & \colhead{} & \colhead{(mag)} & \colhead{(mag)} 
& \colhead{} & \colhead{(mag)} & \colhead{(mag)} 
& \colhead{} & \colhead{(mag)} 
& \colhead{(mag)}}
\startdata
1 & J003659.53-741950.4 & 2001-2008 & 676 & 13.04 & 0.005  & 48 & 14.64 & 0.005 & 23 & 1.61 & 0.007 \\ 
2 & J004441.03-732136.44 & 1997-2008 & 1028 & 14.50 & 0.004 & 82 & 16.10 & 0.006 & 52 & 1.59 & 0.008 \\
3 & J010546.40-714705.27 & 2001-2008 &  690 &  15.00 & 0.006 & 57 & 15.40 & 0.007 & 30 & 0.39 & 0.009 \\ 
4 & J050603.66-690358.9 & 1997-2008  &  537 &  17.56 & 0.016 & 31 & 20.99 & 0.16 & 12 & 3.36 & 0.099 \\ 
5 & J050632.10-714229.8  &1992$-$2000 & 305\tablenotemark{d} & \nodata\tablenotemark{d} & 0.002\tablenotemark{d} & 331\tablenotemark{d} & \nodata\tablenotemark{d} & 0.002\tablenotemark{d} & 298\tablenotemark{d} & \nodata\tablenotemark{d} & 0.003\tablenotemark{d} \\ 
6 & J050713.90-674846.7  & 2001-2008  &  437 &  19.02 & 0.055 & 32 & 21.55 & 0.23 & 10 & 2.48 & 0.14 \\ 
7 & J050835.91-711730.6 & 2001-2008 &  435 &  16.38 & 0.010 & 39 & 18.33 & 0.016 & 13 & 1.96 & 0.019 \\ 
8 & J051110.64-661253.7 & 1993-1999 & 231\tablenotemark{d} & \nodata\tablenotemark{d} & 0.003\tablenotemark{d} & 280\tablenotemark{d} & \nodata\tablenotemark{d} & 0.008\tablenotemark{d} & 225\tablenotemark{d} & \nodata & 0.008\tablenotemark{d} \\ 
9 & J051228.18-690755.8 & 1997-2008 &  855 &  16.13 & 0.008 & 152 & 16.42 & 0.007 & 94 & 0.31 & 0.011 \\ 
10 & J051828.18-680404.0  & 2001-2008 &  305 &  18.28 & 0.031 & 35 & 20.47 & 0.079 & 19 & 2.16 & 0.096 \\ 
11 & J051845.23-700534.5 & 2001-2008 &  493 &  15.80 & 0.007 & 90 & 17.71 & 0.012 & 55 & 1.91 & 0.016 \\ 
12 & J052043.86-692341.0 & 1997-2008 &  957 &  13.98 & 0.005 & 109 & 15.20 & 0.006 & 53 & 1.22 & 0.008 \\ 
13 & J052520.76-705007.5 & 2001-2008 &  435 &  14.19 & 0.006 & 40 & 15.39 & 0.004 & 14 & 1.21 & 0.007 \\ 
14 & J052902.39-661527.7  &1993$-$1999 &  228\tablenotemark{d} & \nodata\tablenotemark{d} & 0.073\tablenotemark{d} & 0\tablenotemark{e} & \nodata & \nodata & 0\tablenotemark{e} & \nodata & \nodata \\ 
15 & J053044.10-714300.6 & 2001-2008 & 426 & 19.92 & 0.13  & 39 & 21.58 & 0.20 & 15 & 1.62 & 0.22 \\
16 & J053250.69-713925.8 & 2001-2008  &  435 &  13.82 & 0.006  & 45 & 15.07 & 0.004 & 17 & 1.25 & 0.007 \\ 
17 & J053525.86-711956.6 & 2001-2008  &  435 &  18.96 & 0.05 & 0\tablenotemark{e} & \nodata & \nodata  & 0\tablenotemark{e} & \nodata & \nodata \\ 
18 & J053632.56-701738.4\tablenotemark{f} & 1997-2008  &  701 &  19.39 & 0.082  & 36 & 21.22 & 0.17 & 11 & 1.83 & 0.19 \\ 
19 & J054054.31-693318.5 & 2001-2008 & 433 & 18.29 & 0.029 & 0\tablenotemark{h} &  \nodata &  \nodata &  \nodata &  \nodata &  \nodata \\
20 & J055311.98-701522.7 & 2001-2008  &  427 &  17.60 & 0.018  & 31 & 19.22 & 0.028 & 13 & 1.66 & 0.031 \\ 
21 & J055825.96-694425.8 & 2001-2008  &  429 &  15.04 & 0.005  & 33 & 15.62 & 0.005 & 13 & 0.58 & 0.007 \\ 
22 & F06111-7023\tablenotemark{g} & 2001-2008  &  795 &  19.89 & 0.13  & 65 & 20.45 & 0.11 & 23 & 0.60 & 0.15 \\ 
\enddata
\tablenotetext{a}{Based on OGLE-II, OGLE-III, and MACHO data.}
\tablenotetext{b}{Average magnitude.}
\tablenotetext{c}{Average uncertainty in an individual observation.}
\tablenotetext{d}{MACHO {\it R$_C$}, {\it V}, and {\it V$-$R$_C$} data; these observations are on the instrumental system, so we have not listed the magnitude values.}
\tablenotetext{e}{No MACHO or OGLE {\it V} data for this object.}
\tablenotetext{f}{\citet{volk11} listed {\it I}=18.1 and {\it V}=18.8 but noted that the \arcsec optical and near-infrared counterpart is $\sim$ 1.5\arcsec away from the SAGE position and may not be the same object.\arcsec  Apparently they indeed did not have the correct optical counterpart.}
\tablenotetext{g}{IRAS ID is listed, since the object is not in the SAGE catalog.}
\tablenotetext{h}{No OGLE {\it V} data for this object.}
\end{deluxetable}

\clearpage

\tablenum{3}
\begin{deluxetable}{rlrcrrrrrrrrrrrrr}
\tablecolumns{16} \tabletypesize{\scriptsize}
\tablecaption{Periodogram Study of the Periodic Light Curves\tablenotemark{a}
\label{periods}}
\rotate
\tabletypesize{\footnotesize} 
\tablewidth{0pt} \tablehead{ 
\colhead{No.} &\colhead{SAGE ID} &\colhead{Years} &\colhead{Filter} & \colhead{P$_1$}&\colhead{A$_1$} &\colhead{$\phi$$_1$\tablenotemark{b}}&\colhead{P$_2$}&
\colhead{A$_2$} &\colhead{$\phi$$_2$\tablenotemark{b}}&\colhead{P$_3$} &\colhead{A$_3$}&\colhead{$\phi$$_3$\tablenotemark{b}}&\colhead{P$_4$} &\colhead{A$_4$}&\colhead{$\phi$$_4$}\tablenotemark{b}& \colhead{P$_2$/P$_1$}} 
\startdata
1 & J003659.53-741950.4 &2001$-$2008 & {\it I} &142.4 & 0.025 & 0.55 & 356.0 & 0.029 & 0.94 & 115.6 & 0.018 & 0.83 & \nodata & \nodata & \nodata & 0.81\tablenotemark{d}\\
2 & J004441.03-732136.44 &1997$-$2008 & {\it I} & 96.3 & 0.041 & 0.22 & 99.6 & 0.023 & 0.01 & 94.4 & 0.023  & 0.34 & \nodata & \nodata & \nodata & 1.03 \\
 & &2001$-$2008\tablenotemark{c} & {\it I} & 96.2 & 0.052 & 0.16 & 92.8 & 0.029 & 0.70 & 100.0 & 0.022  & 0.14 & \nodata & \nodata & \nodata& 0.96 \\
4 & J050603.66-690358.9 &1997$-$2008 & {\it I} &157.2 & 0.105 & 0.04& 137.9 & 0.051 & 0.71 & 239.0 & 0.054 & 0.34 & \nodata  & \nodata & \nodata & 0.88 \\
5 & J050632.10-714229.8 &1992$-$2000 & {\it V} & 48.9 & 0.021 & 0.13 & 53.5 & 0.017 & 0.15 & \nodata & \nodata & \nodata & \nodata & \nodata & \nodata & 1.09\\
8 & J051110.64-661253.7  &1993$-$1999 & {\it R$_C$} &111.8 & 0.015 & 0.56 & 95.7 & 0.013 & 0.58 & \nodata & \nodata & \nodata & \nodata & \nodata & \nodata & 0.86 \\
 &   &1993$-$1999 & {\it V} &96.2 & 0.021 & 0.61 & 112.0 & 0.020 & 0.69 & 500 & 0.017 & 0.61 & \nodata & \nodata & \nodata & 1.16\\
12 & J052043.86-692341.0  &1997$-$2008 & {\it I} &74.2 & 0.0091 & 0.56 & 79.9 & 0.0083 & 0.43 & 57.7 & 0.0073 & 0.26 & 76.9 & 0.0076 & 0.77 & 1.08 \\
16 & J053250.69-713925.8  &2001$-$2008 & {\it I} &90.7 & 0.027 & 0.84 & 85.9 & 0.011 & 0.87 & 80.8 & 0.009 & 0.39 & 117.8 & 0.010 & 0.92 & 0.95\\
19 & J054054.31-693318.5 &2003$-$2008 & {\it I} &144.6 & 0.034 & 0.19 & 130.7 & 0.037 & 0.45 & 168.1 & 0.034 & 0.11 & \nodata & \nodata & \nodata & 0.90\\
\enddata
\tablenotetext{a}{The uncertainties for each object in {\it P}, {\it A}, $\phi$ are approximately as follows:\\
(obj 1) $\pm$0.2 days for P$_1$  and P$_3$ and $\pm$1.2 days for P$_2$, $\pm$0.002 mag, $\pm$0.01, respectively;\\
(obj 2) $\pm$0.1 days, $\pm$0.001 mag, $\pm$0.01, respectively; \\
(obj 4) $\pm$0.2 days, $\pm$0.004 mag, $\pm$0.01, respectively; \\
(obj 5) $\pm$0.1 days, $\pm$0.002 mag, $\pm$0.02, respectively;\\
(obj 8) $\pm$0.4 days, $\pm$0.002 mag, $\pm$0.02, respectively; \\
(obj 12) $\pm$0.1 days, $\pm$0.001 mag, $\pm$0.01, respectively;\\
(obj 16) $\pm$0.1 to $\pm$0.3 days, $\pm$0.001 mag, $\pm$0.03, respectively;\\
(obj 19) $\pm$0.6 days, $\pm$0.003 mag, $\pm$0.02, respectively.\\}
\tablenotetext{b}{The phases are determined based on the epoch of 2,450,000.00.}
\tablenotetext{c}{Based only on the OGLE-III data.}
\tablenotetext{d}{P$_3$/P$_1$, since P$_2$=356 days is suspiciously close to a one year sampling alias.}
\end{deluxetable}

\clearpage

\begin{deluxetable}{rrrrrcrrrl}
\rotate \tablenum{4}
\tablecolumns{10} \tabletypesize{\scriptsize}
\tablecaption{Light Curve Properties and Object Variability
\label{var}} 
\tabletypesize{\footnotesize} 
\tablewidth{0pt} 
\tablehead{
\colhead{No.} 
& \colhead{$\Delta${\it I}\tablenotemark{a}} & \colhead{$<$$\Delta${\it I}$>$\tablenotemark{b}} & \colhead{$\Delta${\it V}\tablenotemark{a}}
& \colhead{$\Delta$({\it V$-$I})\tablenotemark{a}} & \colhead{Color}& \colhead{P} & \colhead{P$_{\rm lit}$\tablenotemark{c}} & \colhead{Ref\tablenotemark{d}} 
& \colhead{Comments\tablenotemark{e}}\\
\colhead{} & \colhead{(mag)} & \colhead{(mag)} & \colhead{(mag)}& \colhead{(mag)} &  \colhead{trend?\tablenotemark{f}} & \colhead{(days)} & \colhead{(days)} & \colhead{} &  \colhead{}
}\startdata
1 & 0.21 & 0.16 & 0.15 & 0.06 & N & 142.4 & 102 & 1   & beat periods (356:, 115 days) \\
2 & 0.22 & 0.15 & 0.40 & 0.18 & Y & 96.3 & 96 & 1,2,3 & beat periods (100, 94 days); LC peaks then declines by 0.08 mag  \\
3 & 0.36 & 0.31 & 0.24 & 0.12 & N & \nodata & \nodata & \nodata  & var. timescale $\sim$10-30 days  \\ 
4 & 0.62 & 0.34 & 0.70 & 0.67 & Y & 157.2 & 159 & 4 & beat periods (138, 239 days)  \\ 
5 & (0.15)\tablenotemark{g} & \nodata & 0.22 & (0.16)\tablenotemark{g} & Y & 48.9 & 49 & 5  & beat period (54 days)  \\
6 & 0.50 & 0.32 & 0.85 & 0.90 & Y & \nodata & \nodata & \nodata & brightness inc. 0.18 mag ({\it I}) 2001-2008; short timescale var.\tablenotemark{h} \\ 
7 & 0.44 & 0.35 & 0.22 & 0.07 & Y & \nodata & 363, 496 & 6,7 & timescale of var. 8$-$20 days; fainter in 2002 by $\sim$0.15 mag \\ 
8 & (0.14)\tablenotemark{g} & \nodata & 0.20  & (0.19)\tablenotemark{g} & Y & 111.8, 96.0 & 96,112 & 2 & these two periods beat against each other \\ 
9 & 0.40 & 0.31 & 0.28 & 0.09 & N & \nodata & \nodata& \nodata  & short timescale var.\tablenotemark{h}  \\ 
10 & 0.35 & 0.20 & 0.4 & 0.5 & Y & \nodata & \nodata & \nodata & brightness dec. 0.04 mag ({\it I}) 2006-2008; short timescale var.\tablenotemark{h}  \\ 
11 & 0.25 & 0.21 & 0.20 & 0.07 & N & \nodata & 39,41 & 1,7  & timescale of var. $\sim$10-35 days  \\ 
12 & 0.09 & 0.08 & 0.14 & 0.07 & Y & 74.2 & 74 & 1,6 & beat periods (80, 77, 58 days)\\
13 & 0.10 & 0.08 & 0.10 & 0.04 & Y & (34.5:) & 34,43 & 1,6 & brightness dec. 0.02 mag ({\it I}) 2001-2008; var. timescale $\sim$10-60 days  \\ 
14 & (0.34)\tablenotemark{g} & \nodata & \nodata & \nodata & \nodata & \nodata & \nodata & \nodata & var. timescale $\sim$10-20 days   \\ 
15 & 0.92 & 0.71 & 1.1: & 1.0: &Y & \nodata & \nodata & \nodata & short timescale var.\tablenotemark{h} \\
16 & 0.14 & 0.10 & 0.23 & 0.13 & Y & 90.7 & 91$-$93 & 1,6,7  & beat periods (86, 81, 118 days) \\
17 & 0.38 & 0.33 & \nodata & \nodata & \nodata & \nodata & \nodata & \nodata & brightness inc. 0.19 mag, dec. 0.09 mag ({\it I}); short timescale var.\tablenotemark{h} \\ 
18 & 0.69 & 0.49 & 0.7 & 0.6 & Y & \nodata & \nodata & \nodata  & brightness lower by 0.10 mag ({\it I}) 1997-1998; short timescale var.\tablenotemark{h}  \\ 
19 & 0.29 & 0.24 & \nodata & \nodata & \nodata & 144.6 & 5.2 & 6  & beat periods (131, 168 days); brightness inc. by 0.5 mag in 2001 \\
20 & 0.21 & 0.17 & 0.14 & 0.08 & N & \nodata & \nodata & \nodata &  brightness inc. $\sim$0.08 mag (I); short timescale var.\tablenotemark{h}  \\ 
21 & 0.32 & 0.25 & 0.20 & 0.03 & Y & \nodata & \nodata & \nodata & brightness dec. 0.10 mag ({\it I}) 2001-2003; var. timescale $\sim$10 days  \\ 
22 & 1.03 & 0.82 & 0.3: & 0.5: & Y & \nodata & \nodata & \nodata  & brightness inc. 0.13 mag ({\it I}) 2001-2008; short timescale var. $\le$8 days \\ 
\enddata
\tablecomments{Uncertain values are followed by colons.}
\tablenotetext{a}{Maximum variation in a season.}  
\tablenotetext{b}{Average variation in a season.}
\tablenotetext{c}{Published period values.}
\tablenotetext{d}{References for the published period values: (1) \citet{sos11}; (2) \citet{smedt12}; (3) \citet{kam14}; (4) \citet{ita04}; (5) \citet{aarle11}; (6) \citet{fraser08}; (7) \citet{spano11}.}
\tablenotetext{e}{Abbreviations are defined as follows: var. = variability, dec. = decreasing, inc. = increasing.}
\tablenotetext{f}{{V}$-$({V$-$I)} or {V}$-$({V$-$R$^{\prime}_C$}) color trends (redder when fainter)? $-$ Y=yes, N=no.}
\tablenotetext{g}{Values in parentheses are MACHO {\rm R}$^{\prime}_C$ and ({\rm V}$^{\prime}$$-${\rm R}$^{\prime}_C$) values, rather than OGLE {\it I} and {\it V$-$I} values.}
\tablenotetext{h}{Short timescale variability of a few days or less.}
\end{deluxetable}

\clearpage

\tablenum{5}
\begin{deluxetable}{rlrrrrrrrcl}
\rotate 
\tablecolumns{10} \tabletypesize{\scriptsize}
\tablewidth{8.0truein}
\tablecaption{Table of Results
\label{results}} 
\tablehead{
\colhead{No.} & \colhead{SAGE ID} & \colhead{$<${\it I}$>$} & \colhead{$<${\it V$-$I}$>$} 
& \colhead{$\Delta${\it I}\tablenotemark{a}} & \colhead{$\Delta${\it V}\tablenotemark{a,b}} & \colhead{P\tablenotemark{c}} & \colhead{SpT\tablenotemark{d}} & \colhead{T$_{\rm eff}$\tablenotemark{e}} 
& \colhead{21 $\mu$m} &  \colhead{Classification}\\
\colhead{} & \colhead{} & \colhead{(mag)} & \colhead{(mag)} 
& \colhead{(mag)} & \colhead{(mag)} & \colhead{(days)} & \colhead{} & \colhead{(K)} 
& \colhead{Strength\tablenotemark{f}} &  \colhead{}
}\startdata
1 & J003659.53-741950.4 & 13.04 & 1.61  & 0.21 & 0.15\tablenotemark{g} & 142 & \nodata & 5500:\tablenotemark{l}  & N & PPN: \\
2 & J004441.03-732136.44 & 14.50 & 1.59  & 0.22 & 0.40\tablenotemark{g}& 96 & \nodata & 6250 & Y & PPN  \\
3 & J010546.40-714705.27 & 15.00 & 0.39 & 0.36 & \nodata &  m & \nodata & \nodata & Q & likely hot (A, late-B) PPN  \\ 
4 & J050603.66-690358.9 & 17.56 & 3.36  & 0.62 & 0.70\tablenotemark{g,h} & 157 & \nodata & \nodata & Q & PPN \\ 
5 & J050632.10-714229.8 & 13.78\tablenotemark{i} & 0.52\tablenotemark{i}  & (0.15)\tablenotemark{j} & 0.22 & 49 & A3Iab, F0Ibp(e) & 6750 & Un & PPN  \\ 
6 & J050713.90-674846.7  & 19.02 & 2.48  & 0.50 & \nodata & s & \nodata & \nodata & W & likely hot PPN/young PN  \\ 
7 & J050835.91-711730.6 & 16.38 & 1.96  & 0.44 & \nodata & m & late-B$-$early-G  & \nodata & Y & likely hot (A, late-B) PPN  \\
8 & J051110.64-661253.7 & 14.54\tablenotemark{i} & 2.09\tablenotemark{i}  & (0.14)\tablenotemark{j} & 0.20 & 112, 96 & F3II(e), G5I & 6750 & Y & PPN  \\ 
9 & J051228.18-690755.8 & 16.13 & 0.31 & 0.40 & \nodata & s & B0.5I(e) & \nodata & N & hot PPN/young PN \\ 
10 & J051828.18-680404.0  & 18.28 & 2.16  & 0.35 & \nodata & s & \nodata & \nodata & Y & likely hot PPN/young PN \\ 
11 & J051845.23-700534.5 & 15.80  & 1.91  & 0.25 & \nodata& m & H$\alpha$ em & \nodata & Y & likely hot (A, late-B) PPN  \\ 
12 & J052043.86-692341.0 & 13.98 & 1.22 & 0.09 & 0.14\tablenotemark{g} & 74 & F5Ib(e), F8I(e), G5I  & 5750 & Y & PPN  \\
13 & J052520.76-705007.5 & 14.19 & 1.21  & 0.10  & \nodata & m & A1Ia/F2-5Ia, F3I  & 6750 & Y & PPN \\ 
14 & J052902.39-661527.7  & 17.35\tablenotemark{i} & 4.0\tablenotemark{i}  & 0.34\tablenotemark{j} & \nodata& m & \nodata & \nodata & W & likely hot (A, late-B) PPN  \\ 
15 & J053044.10-714300.6 & 19.92 & 1.62  & 0.92  & \nodata & s & \nodata & \nodata  & N &likely hot PPN/young PN \\
16 & J053250.69-713925.8 & 13.82 & 1.25  & 0.14 & 0.23\tablenotemark{g} & 91 & F6Ia   & 5500  & Un & PPN \\
17 & J053525.86-711956.6 & 18.96 & \nodata  & 0.38 & \nodata & s & \nodata & \nodata  & Q & likely hot PPN/young PN  \\ 
18 & J053632.56-701738.4 & 19.39 & 1.83  & 0.69 & \nodata& s & \nodata & \nodata & N (SiC?) & likely hot PPN/young PN  \\ 
19 & J054054.31-693318.5 & 18.29 & 3.47  & 0.29\tablenotemark{m} & \nodata & 145 & \nodata  & \nodata  & N & PPN, with RCB properties \\
20 & J055311.98-701522.7 & 17.60 & 1.66  & 0.21 & \nodata& s & \nodata & \nodata & N (SiC?) & likely hot PPN/young PN \\ 
21 & J055825.96-694425.8  & 15.04 & 0.58  & 0.32 & \nodata& s & F6Ia  & \nodata & N &likely hot PPN/young PN  \\ 
22 & F06111-7023\tablenotemark{k} & 19.89 & 0.60  & 1.03 & \nodata& s & \nodata & \nodata & Q & likely hot PPN/young PN  \\ 
\enddata
\tablenotetext{a}{Maximum variation in a season.}
\tablenotetext{b}{We have only listed $\Delta${\it V} for the objects that show periodic variability.}  
\tablenotetext{c}{Periodicity: s = short timescale variation of $\le$ 10 days, and usually a few days or less, no period determined; m = medium timescale variation of 10$-$35 days, no period determined.}
\tablenotetext{d}{Optical spectral types from \citet{aarle11} (nos. 5, 8, 11, 12, 13, 16), this paper (nos. 9, 12), \citet{mat14} (nos. 7, 13, 21), and \citet{haj15} (nos. 8, 12, 13).}
\tablenotetext{e}{Temperatures from \citet{smedt12} (no. 2), \citet{haj15} (no. 8), and  \citet{aarle13} (nos. 5, 12, 16). }
\tablenotetext{f}{21 $\mu$m presence and strength: Y = strong, W = weak, Q = very weak or questionable if present, N = not present; based on study by \citet{volk11}, except for three additional ones that we classified based on spectra presented by \citet{sloan14} (objects 1, 15, 19). Un = unknown; no Spitzer SAGE spectra of this object.}
\tablenotetext{g}{These are minimum values since there are many fewer OGLE {\it V} than {\it I} observations (in contrast to the MACHO observations, in which there were the same number of {\rm V}$^{\prime}$ and {\rm R}$^{\prime}_C$ observations).}
\tablenotetext{h}{Object 3 is very faint in {\it V} with large uncertainties.  To get a more realistic estimate of the maximum $\Delta${\it V} in a season, we used only values of V with $\sigma$$_V$ $\le$ 0.15 mag.}
\tablenotetext{i}{No OGLE data; values from MPSC catalog \citet{zar97}.}
\tablenotetext{j}{These are $\Delta${\rm R}$^{\prime}_C$ values from MACHO.}
\tablenotetext{k}{IRAS ID is listed, since the object is not in the SAGE catalog.}
\tablenotetext{l}{Temperature is uncertain, and is based on similarity of the spectrum to that of MWG PPN IRAS Z02229+6208.}
\tablenotetext{m}{Omitting seasons 1 and 2 (2001$-$2002) and the first four observations of season 3.}
\end{deluxetable}

\clearpage

\tablenum{6}
\begin{deluxetable}{rlrcccrr}
\rotate \tablenum{6}
\tablecolumns{7} \tabletypesize{\scriptsize}
\tablecaption{Temperature Variations for Those With Contemporaneous High-Resolution Spectra
and Light Curves Observations
\label{Tvar}} 
\tablehead{
\colhead{No.} & \colhead{SAGE ID} & \colhead{T$_{\rm eff}$} & \colhead{$\Delta$({\it V$-$I})$_C$\tablenotemark{a,b}} & \colhead{$\Delta$({\it V$-$I})$_J$\tablenotemark{a,c}} 
& \colhead{LC Phase} & \colhead{$\Delta$T$_{\rm eff}$\tablenotemark{a}} & \colhead{T$_{\rm eff}$~Range\tablenotemark{a}} 
\\
\colhead{} & \colhead{} & \colhead{(K)} & \colhead{(mag)} 
& \colhead{(mag)} & \colhead{} & \colhead{(K)} & \colhead{(K)} 
}\startdata
2 & J004441.03-732136.44 & 6250 & 0.18 & 0.23 & mid-level & 1100 & 5750$-$6850  \\
12 & J052043.86-692341.0 & 5750 & 0.07 & 0.08 & minimum & 350 & 5750$-$6100  \\
16 & J053250.69-713925.8 & 5500 & 0.13 & 0.16 & slightly below mid-level  & 430 & 5320$-$5750  \\
\enddata
\tablenotetext{a}{Maximum variation in a season.}
\tablenotetext{b}{Observed on the Cousins system}
\tablenotetext{c}{Transformed to the Johnson system.}
\end{deluxetable}

\clearpage

\clearpage

\end{document}